\documentclass[conference]{IEEEtran}
\ifCLASSINFOpdf
\else
\fi
\usepackage{booktabs}
\usepackage{tikz}
\usepackage{amsmath}
\usepackage{amssymb}
\usepackage{comment}
\usepackage{subcaption}
\usepackage{multirow}
\usepackage{xspace}
\usepackage{makecell}
\usepackage{xcolor}
\usepackage{pifont} 
\usepackage{url}
\usepackage[normalem]{ulem} 
\newcommand{\yazan}[1]{\vspace{0.3cm}\noindent\textcolor{blue}{\textbf{[Yazan]:} #1}}

\usepackage{todonotes}
\newcommand{\sys}{\textsc{CallShield}\xspace}
\newcommand{\heading}[1]{{\noindent\bf{#1}}~}

\begin{document}
%


\title{\sys: Secure Caller Authentication over Real-Time Audio Channels}


\author{\IEEEauthorblockN{Mouna Rabhi, Yazan Boshmaf, Mashael Alsabah, Shammur Chowdhury, Mohamed Hefeeda, and Issa Khalil}
	\IEEEauthorblockA{\em Qatar Computing Research Institute, HBKU}
}

\maketitle

\begin{abstract}
We present \sys, the first caller identity authentication system that operates entirely at the audio layer, without relying on speech transcription, internet connectivity, or trusted infrastructure. \sys introduces a real-time neural watermarking technique that enables per-bit embedding and recovery within 40-millisecond frames of live 8 kHz speech. This capability allows \sys to transform the real-time audio channel into a noisy serial communication medium. To ensure reliable data transmission, \sys implements a low-bitrate data link protocol that provides basic frame synchronization along with error detection, correction, and recovery. For caller authentication, \sys adopts a secure and lightweight symmetric-key protocol that relies on pairwise shared secrets among trusted contacts. The system completes the full authentication process in an average of 63 seconds, including up to three retransmission attempts, making it suitable for real-time deployment. Extensive experiments under realistic telephony conditions demonstrate that \sys achieves an overall authentication success rates exceeding 99.2\% on clean audio and over 95\% under common distortions, aided by selective retransmission of failed messages. Additionally, \sys maintains high audio quality, achieving PESQ scores above 4.2 and STOI scores above 0.94 on clean speech, and exhibits robustness across a wide range of channel distortions, validating its practical viability for secure, real-time caller authentication.
\end{abstract}


%
\IEEEpeerreviewmaketitle

\section{Introduction}
\label{sec:intro}

\heading{Motivation.} Phone calls remain a trusted and personal mode of everyday communication, making them an increasingly attractive target for financial fraud.
Scam calls, in particular, represent a growing global security threat in which attackers impersonate banks, government agencies, or service providers to manipulate victims into disclosing sensitive information. These attacks rely not only on spoofed caller ID information (i.e., caller phone number or name), but also on increasingly sophisticated social engineering tactics that exploit urgency, authority, or fear to build credibility and pressure victims~\cite{sahin2017sok}. Even well-informed individuals can be deceived, especially when the caller fabricates plausible scenarios involving fraud alerts, account verification, or overdue payments. A recent study by the Consumer Federation of America reports that Americans lost \$16 billion to fraud in 2024, with nearly half of the victims contacted via phone calls or text messages~\cite{consumerfed}.

\heading{Problem.} Despite significant research and commercial deployment, existing defenses against scam calls remain reactive, fragmented, and ultimately ineffective. Caller identity filters and blacklists~\cite{miramirkhani2016dial, pandit2018towards} are easily bypassed through caller ID spoofing or the use of frequently changing phone numbers. Approaches based on call content post-hoc analysis, such as those using Natural Language Processing (NLP)~\cite{zhao2018detecting, nicholas2024scamdetector, shen2025warned}, rely on full speech transcription, require internet connectivity for cloud-based services, and remain inherently reactive, attempting to infer trustworthiness only after the call has been received. Network-level caller identity authentication frameworks, such as STIR/SHAKEN~\cite{transnexus2022stir}, depend on VoIP infrastructure and have not been universally adopted, particularly in cross-network contexts or legacy telephony environments~\cite{FCC2025CallerID}.

\heading{Proposed Solution.} Unlike prior work, we address a more fundamental challenge in detecting scam calls: authenticating and verifying the caller's identity in real time without requiring modifications to the underlying telephony infrastructure. We introduce \sys, a novel, low-latency, symmetric-key based caller authentication system that runs on network endpoints (e.g., mobile phones or VoIP clients) and operates entirely at the real-time audio channel of live voice calls. \sys embeds authentication messages directly into the speech signal using neural audio watermarking. These messages are embedded and decoded locally at the endpoints, without relying on cloud services or speech transcription. By operating solely on narrowband audio signals, \sys maintains compatibility with standard PSTN, mobile, and VoIP telephony networks, and avoids the need for widespread network-level deployment. Authentication is performed using pairwise symmetric keys shared offline between individuals and their trusted contacts, enabling secure, real-time identity authentication and verification during live calls.

\heading{Design.} \sys offers three services that collectively enable real-time authentication of the caller's identity:


\textit{(1) Real-time audio watermarking:} State-of-the-art (SOTA) watermarking techniques, such as AudioSeal~\cite{roman2024proactive}, SilentCipher~\cite{singh2024silentcipher}, WMCodec~\cite{zhou2025wmcodec}, and Timbre~\cite{liu2023detecting} rely on long audio frames ($\ge 1$ second) and high sampling rates ($>8$kHz), making them unsuitable for live, low-bitrate speech watermarking. To overcome these limitations, \sys introduces a Real-Time Audio Watermarking (RTAW) model that builds on AudioSeal's neural generator-detector architecture and integrates a new attention-based watermark embedding module. This enhancement leads to better watermark recovery under various distortions, especially in shorter audio frames. To enable real-time operation, \sys employs an RTAW model trained on the shortest audio frame that supports the recovery of one or more watermarked bits per frame. Specifically, the model embeds one bit per 40-millisecond frame at an 8kHz sampling rate, resulting in an endpoint-to-endpoint latency of less than 200 milliseconds.

\textit{(2) Reliable data-over-audio communication:} With RTAW, \sys models the real-time audio channel of a live voice call as a noisy serial communication medium characterized by binary symmetric errors (i.e., random single- or multi-bit flips). As there are no standardized protocols for data transmission over watermarked real-time audio, \sys introduces a lightweight data link protocol designed specifically for this setting. The protocol supports low-bitrates (25~bps), provides frame-level synchronization, and incorporates bit-level error detection, correction, and recovery through the use of Bose-Chaudhuri-Hocquenghem (BCH) codes and a stop-and-wait Automatic Repeat reQuest (ARQ) mechanism.

\textit{(3) Lightweight endpoint authentication:} Using symmetric-key cryptography, \sys implements a four-message, secure challenge-response protocol for endpoint authentication over its data-over-audio communication channel. The protocol is based on pairwise secrets initially shared offline between trusted entities (e.g., an individual and a family member, financial institution, or government agency). In particular, it uses a Hash-based Message Authentication Code (HMAC) to verify that the calling endpoint (i.e., the caller) possesses the secret key previously shared with the called endpoint (i.e., the callee), thereby authenticating the caller's identity.

\heading{Evaluation.} RTAW demonstrates strong bit-level performance, achieving 98\% bit accuracy on clean speech at maximum watermark strength (\( \alpha = 1.0 \)) \footnote{ \( \alpha \) controls the amplitude of the perturbation added during embedding; higher values increase robustness at the cost of audio quality.}, and maintains above 90\%  under common telephony distortions.  At lower watermark strength, i.e., $\alpha=0.6$, bit accuracy remains robust (above 91\%), with improved  perceptual quality. However, the perfect recovery rate, i.e., full bitstream recovery without any bit errors, is more fragile. For example, the perfect recovery rate is  67\%, in clean conditions. 
To address this, \sys layers a reliable data-over-audio communication protocol atop RTAW, incorporating synchronization, error correction, and bounded retransmission. These mechanisms recover from bit flips and misalignment error. As a result,  \sys achieves an overall authentication success rate of 99.2\% on clean audio and over 95\% under distortions, while maintaining a 0\% false acceptance rate against adversaries lacking the shared key. The full authentication process completes in an average of 63 seconds, making the system practical for real-time deployment

\heading{Contributions.} This paper makes the following contributions:

\begin{itemize}
\item We propose \sys, the first real-time, audio-layer caller authentication system that embeds cryptographically verifiable identity information directly into live speech using neural watermarking.
\item We design and implement a lightweight, frame-level watermarking system that operates on narrowband audio and enables per-bit embedding and detection with low latency with average latencies of 12.1 ms (embedding) and 4.2 ms (decoding), while preserving perceptual quality (PESQ$>$4.2, STOI $>$0.96).
\item We propose a reliable data link protocol over audio tailored for real-time watermarking audio. It models the watermarking system as a binary symmetric channel and introduces frame synchronization, BCG error correction, and retransmission to ensure robust message delivery.
\item We build a lightweight, symmetric-key challenge-response protocol for authentication over audio, enabling real-time identity verification.
\item We conduct a comprehensive evaluation under realistic call conditions, covering noise, delay, filtering, and compression, and demonstrate that \sys achieves 99.2\% authentication success rates  on clean speech and over 95\% on common distortions, while maintaining a 0\% false acceptance rate against unauthorized attackers.
\end{itemize}


\section{Background and Related Work}
\label{sec:background}
Modern telecommunication networks -- including landlines, cellular, and VoIP systems -- offer varying levels of caller ID security, but remain vulnerable to spoofing. While recent work has explored audio-layer approaches for caller authentication, many fall short in real-time or offline scenarios, highlighting the need for more practical and robust solutions.

\heading{Network-level Authentication (STIR/SHAKEN).} To mitigate VoIP spoofing, the FCC mandated STIR/SHAKEN, a framework that signs SIP headers to verify the originating provider~\cite{fcc2025combating}. It assigns one of three attestation levels (A/B/C) based on the provider's confidence in the caller's authority~\cite{sherman2021characterizing}, but does not authenticate the caller's identity.
STIR/SHAKEN has key limitations. It depends on a U.S.-centric PKI that does not extend globally~\cite{yu2021analysis}, supports only SIP/IP networks (excluding SS7-based systems)\cite{transnexus2022reply}, and often loses signature data across non-IP segments\cite{transnexus2022stir}. Attestation levels are opaque to users, and spoofed calls persist—25\% of B-level and 33\% of C-level calls are still robocalls~\cite{transnexus2022robocalls}. Despite IETF standardization~\cite{ietf2023secure}, the system has not meaningfully curbed spoofing, and alternatives remain underexplored.

\heading{Blacklist-based Detection.} 
Blacklist-based detection has emerged as a practical defense against phone-based scams, primarily by identifying and blocking known fraudulent numbers.
 Miramirkhani \emph{et al.}\cite{miramirkhani2016dial} analyzed technical support scams and built a system that detects scam pages via behavioral cues, enabling the identification and blocking of fraudulent numbers. The work emphasizes user interaction as a persistent attack vector and advocates for proactive, behavior-driven defenses. Pandit \emph{et al.}\cite{pandit2018towards} showed that blacklists built from complaints, crowd-sourced reports, and honeypots can preemptively block many scam calls. However, blacklisting remains reactive and is easily evaded through caller ID spoofing or rapid number rotation.

\heading{Content-based Detection.} Zhao \emph{et al}.~\cite{zhao2018detecting} proposed a content-based approach that analyzes the semantic content of calls using NLP techniques rather than relying on metadata such as phone numbers. Their method extracted linguistic features from user-reported fraud descriptions. However, the reliance on basic NLP tools limited its adaptability in rapidly evolving fraud scenarios. Similarly, Bajaj  \emph{et al.}~\cite{bajaj2019fraud} used transcribed call conversations as input features and employed various machine learning classifiers to distinguish between legitimate and fraudulent interactions. While these approaches offer promise, they often require accurate and timely transcription pipelines and suffer from latency and generalization issues in real-time applications. Moreover, transcribing the entirety of a phone call raises significant privacy concerns, especially when dealing with sensitive or legally protected information.

\heading{Watermark-based Detection.} Audio watermarking has emerged as a promising approach for copyright protection and content provenance. Traditional methods, such as echo hiding and spread spectrum, modify signal features with hand-crafted rules to embed information. While effective under mild conditions, these techniques often struggle against modern distortions like compression and generative resynthesis. 
Neural audio watermarking has emerged as a robust alternative, leveraging end-to-end learning to embed and extract binary messages resilient to noise, filtering, and re-encoding. Recent work focuses on audio authentication, especially for detecting synthetic speech. AudioSeal~\cite{roman2024proactive} embeds 16-bit payloads using an EnCodec-style architecture for offline detection and localization. SilentCipher~\cite{singh2024silentcipher} uses psychoacoustic masking and compression-aware training for high-capacity, imperceptible watermarking. WMCodec~\cite{zhou2025wmcodec} integrates watermarking into neural codecs, while Timbre watermarking~\cite{liu2023detecting} targets speaker-specific features to resist voice cloning. However, these methods are unsuitable for real-time applications because they operate offline, requiring longer audio segments and higher audio sampling rates (see Table \ref{tab:watermark-comparison} for details).

\heading{Real-time Detection.}
Recent work leverages large language models (LLMs) for real-time scam call detection. ScamDetector~\cite{nicholas2024scamdetector} fine-tunes LLMs (e.g., GPT-2, LLaMA-3) on public and synthetic scam datasets tailored to local languages, analyzing the semantics of live calls to detect scams from unknown numbers. While effective, its reliance on synthetic data may limit generalization to real-world scams.
Singh~\emph{et al.}\cite{singh2025advanced} improve adaptability using Retrieval-Augmented Generation (RAG) to verify whether callers solicit sensitive information during live calls. However, the system introduces latency and privacy risks by processing full conversations via centralized LLMs.
Similarly, Shen\emph{et al.}~\cite{shen2025warned} use prompt-based LLMs to detect scam patterns in transcribed voice streams and issue real-time warnings. Though effective in simulation, the approach faces limitations including prompt sensitivity, privacy concerns, and susceptibility to adversarial manipulation.

\section{Threat Model}
\label{sec:threat-model}

The adversary's objective is to deceive the receiver into believing the call originates from a trusted entity  (e.g., a bank or trusted contact), thereby establishing unwarranted trust. A successful impersonation enables the attacker to bypass authentication, mislead the receiver, and elicit confidential or security-critical responses.

\heading{Active Attacker.} Our threat model is inspired by the most common real-work scenario currently used to deceive victims. The adversary seeks to impersonate a trusted caller,such as a government agency, financial institution, or a personal contact, by spoofing the caller ID and delivering deceptive audio content. The attacker's goal is to exploit the perceived legitimacy of the call to manipulate the victim into disclosing sensitive information (e.g., passwords, banking credentials, one-time verification codes) or performing security-critical actions such as approving financial transactions. The attacker may use various tactics to carry out the attack, including:

\begin{itemize}
    \item Caller ID Spoofing: The attacker can manipulate SIP or SS7 metadata to present a falsified caller ID.
    
    \item Audio Injection: The attacker can inject pre-recorded or synthetically generated audio into the call to mimic the speech of a legitimate sender.
    
    \item Replay Attacks: The attacker may replay audio from a prior legitimate call that contains a valid watermark, attempting to reuse previously authenticated content.
    
    \item Full Device Control: The attacker fully controls the originating device and can install arbitrary software, manipulate audio pipelines,  configure codecs and gain levels.
    
    \item Adversarial Watermarking: The attacker may attempt to mimic the behavior of \sys and inject adversarial watermarks to evade detection or spoof authentication.
\end{itemize}


\heading{Attacker limitations.} The adversary is assumed not to possess the cryptographic secrets used to embed or verify watermarks in \sys. The security guarantees of the system rely on the secrecy of these keys. We assume that the adversary operates over traditional cellular or VoIP networks but has no access to or control over the underlying telephony infrastructure. That is, the attacker cannot compromise telecom carriers, collude with network insiders, or hijack in-progress calls. The communication channel is assumed to faithfully carry audio end-to-end, though signal distortions due to compression, jitter, or echo may naturally occur.

\heading{Users.} We also assume that users employ devices equipped with \sys, our proposed authentication system. A shared cryptographic key is established out of band prior to the call, for example, during user registration with a bank or network provider, or through a trusted social contact. This assumption is aligned with standard practices in online banking and prior secure voice communication systems that rely on out-of-band credential exchange~\cite{onespan_oob_authentication}. 

\section{System Model and Problem Statement}
\subsection{System Model}
\begin{figure}
    \centering
    \includegraphics[width=\linewidth]{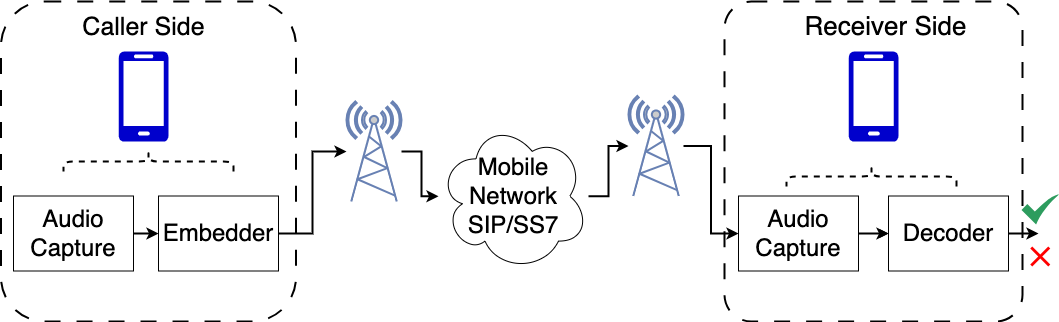}
    \caption{An overview of the proposed end-to-end system}
    \label{fig:system-design-overview}
\end{figure}

We consider a two-party communication setting comprising a \emph{caller} \( c \in C \) and a \emph{receiver} \( r \), engaged in a real-time audio exchange over a standard cellular voice channel (e.g., PSTN, 2G/3G/4G networks). The interaction occurs entirely over the telephony path, without reliance on VoIP, internet connectivity, or auxiliary metadata channels. The goal of the system is to enable a receiver \(r \in  R \) to verify, in real time and using only the received audio signal, that the call originates from a legitimate and trusted caller \( c \in C \).
Figure~\ref{fig:system-design-overview} illustrates the overall communication architecture of \sys. On the caller's side, live speech is captured from the microphone and processed by a lightweight neural watermark embedder, which encodes an authentication message into the outgoing audio stream. The watermarked audio is then transmitted through the cellular network using the standard voice channel, undergoing typical channel degradations such as compression, jitter, and noise.

On the receiver's side, the audio is captured from the device speaker and fed into a real-time watermark decoder, which extracts the embedded bits on a frame-by-frame basis. The extracted message is then passed to the system logic for verification and state progression. 

Each trusted caller \( c \) shares a symmetric key \( K \) with the receiver \( r \). This key is established out-of-band prior to the call and is used to generate cryptographic authentication tokens embedded in the speech signal. 
\subsection{Problem Statement}
The problem addressed in this work is real-time caller authentication over standard telephony audio, without relying on transcription, internet access, or third-party infrastructure. The receiver must be able to verify, during an ongoing live call, that  the receiver is from a pre-trusted caller.
\begin{itemize}
    \item \textbf{Real-Time Operation.}  The system must operate within the strict latency constraints of a live call. Specifically, delays exceeding 200 ms can cause noticeable echo, overlap in turn-taking, and user frustration~\cite{ITU-G114}. Essentially, audio must be watermarked and verified on the fly, without perceptible delays. To preserve conversational fluidity, end-to-end latency must remain below 200 ms, as higher delay degrades call quality. \sys  introduces a real-time audio watermarking technique that processes 40~ms audio frames, enabling authentication without introduce perceptible lag.
    \item \textbf{Imperceptibility.}  Retaining natural audio quality is critical, as human listeners are highly sensitive to even minor speech distortions. This significantly limits the energy and density of any watermark signal.  To preserve imperceptibility, we constrain the watermark strength so that it introduces only minimal, inaudible changes to the original speech signal.
    \item \textbf{Robustness.} Audio transmitted over cellular networks is often degraded by  jitter, noise, and other distortions. These distortions can disrupt embedded signals and lead to decoding errors.   \sys train the watermarking model with diverse simulated channel conditions and use error-correcting codes to maintain high recovery rates under distortion.
    \item \textbf{Authenticity.} Only trusted callers should be able to generate valid authentication signals. \sys embed cryptographically derived responses tied to a shared key using a challenge–response protocol, ensuring that only legitimate senders can produce valid response. 
\end{itemize}

\section{Proposed Solution}
\subsection{Overview}

\sys introduces a novel, low-latency, symmetric key-based caller authentication system in real-time, without requiring modifications to the underlying telephony structure. It embeds authentication messages directly in the live audio speech using a real-time watermarking technique. 
The \sys comprises three key components:

\begin{enumerate}
\item \textbf{Real-Time Audio Watermarking (RTAW):} A lightweight, frame-synchronous watermarking method that causally embeds one bit of information into each audio frame (or segment) of duration $d=40$ms. This design enables real-time inference, achieving high transparency and robustness under channel distortions.

\item \textbf{Data Link Protocol Over Audio:} \sys models RTAW as a noisy serial communication medium with binary symmetric errors and implements a lightweight data link protocol for reliable message delivery, offering synchronization, error detection and correction, and retransmission. Each message frame includes a synchronization preamble, a BCH-encoded payload,  enabling robust framing, alignment, and error correction under real-time audio conditions. 

\item \textbf{Endpoint Authentication Protocol:} A lightweight, symmetric-key-based challenge–response authentication protocol built atop the data link protocol communication channel. The protocol uses shared secret keys initially shared offline between trusted contacts. More specifically, it uses a  Hash-based Message Authentication Code (HMAC) to verify that the caller possesses the secret key previously shared with the receiver, thereby authenticating the caller’s identity.
\end{enumerate}

\begin{figure}[]
    \centering
    \includegraphics[width=0.95\linewidth]{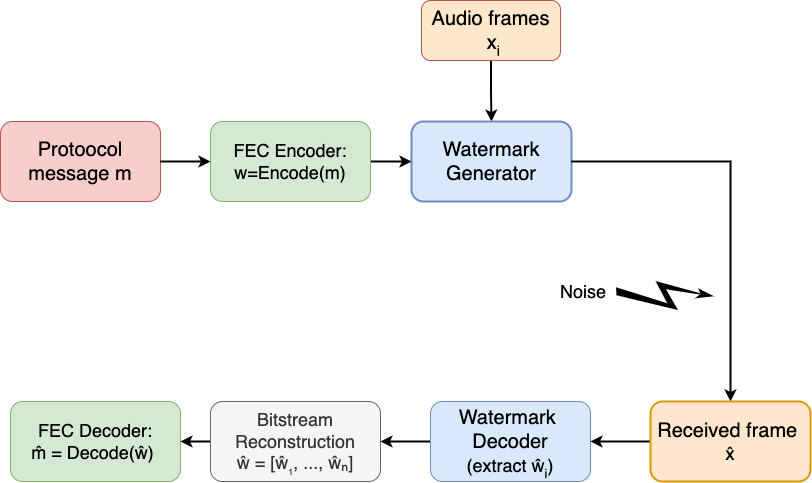}
    \caption{Overview of how messages are embedded and extracted using the watermark generator and decoder.} 
    \label{fig:watermarking-diagram}
\end{figure}

\noindent \sys is modular by design, with each component collectively contributing to ensure robust, real-time, and reliable system operation. Figure~\ref{fig:watermarking-diagram} illustrates how protocol messages are embedded, transmitted, and decoded in real-time.

\subsection{Real-time Audio Watermarking}
\label{sec:system-design-watermark-generator}
\heading{Limitations of the current approaches}
Existing methods such as Audioseal~\cite{roman2024proactive}, SilentCipher~\cite{singh2024silentcipher}, WMCodec~\cite{zhou2025wmcodec}, and Timbre~\cite{liu2023detecting} demonstrate strong performance in offline or high-bitrate scenarios, but fail to meet the strict requirements of real-time and low-bitrate telephony.
AudioSeal~\cite{roman2024proactive} introduces fine-grained, sample-level watermarking for detecting synthetic speech, however operates on 1 second 16kHz clips and requires offline processing. 
SilentCipher \cite{singh2024silentcipher} embeds robust, high-capacity watermarks using psychoacoustic masking and SDR constraints, however the model also operates on 44.1 kHz sampling rate with an audio segment of 1 second.
WMCodec~\cite{zhou2025wmcodec} integrates watermarking into neural codecs to improve robustness under compression, yet its codec-level operation and dependence on longer segments make it incompatible with narrowband telephony signals. Similarly, Timbre watermarking~\cite{liu2023detecting} improves resistance to voice cloning by targeting speaker timbre features, but again assumes wideband audio and offline inference.

All existing systems (see Table~\ref{tab:watermark-comparison}) typically require access to the entire or a large segment of audio stream ($\geq 1$ second), operates at high sampling rates, and lack frame-level processing capabilities. None of these current approaches support frame-by-frame watermarking at an 8 kHz sampling rate with end-to-end latency under 200 ms. These limitations highlight the need for a new watermarking solution specifically designed for frame-synchronous, real-time communication over narrow bandwidth-constrained voice channels.

\begin{table}[t]
\centering
\caption{Comparison of prior audio watermarking systems.}
\label{tab:watermark-comparison}
\begin{tabular}{lcccc}
\toprule
\textbf{System} & \textbf{Real-Time} & \makecell{\textbf{Sampling} \\ \textbf{Rate}} & \makecell{\textbf{Audio} \\ \textbf{Length}} & \makecell{\textbf{Capacity} \\ \textbf{(bit/s)}} \\
\midrule
AudioSeal~\cite{roman2024proactive} & \ding{55} & 16\,kHz & 1\,s & 16 \\
SilentCipher~\cite{singh2024silentcipher} & \ding{55} & 44.1\,kHz & 1\,s & 40 \\
WMCodec~\cite{zhou2025wmcodec} & \ding{55} & 12\,kHz & 1\,s & 13.3 \\
Timbre~\cite{liu2023detecting} & \ding{55} & 16\,kHz & $>$1\,s & 100 \\
RTAW (Ours) & \ding{51} & \textbf{8\,kHz} & \textbf{40\,ms} & \textbf{25} \\
\bottomrule
\end{tabular}
\end{table}

\begin{figure*}[]
    \centering
    \includegraphics[width=0.83\linewidth]{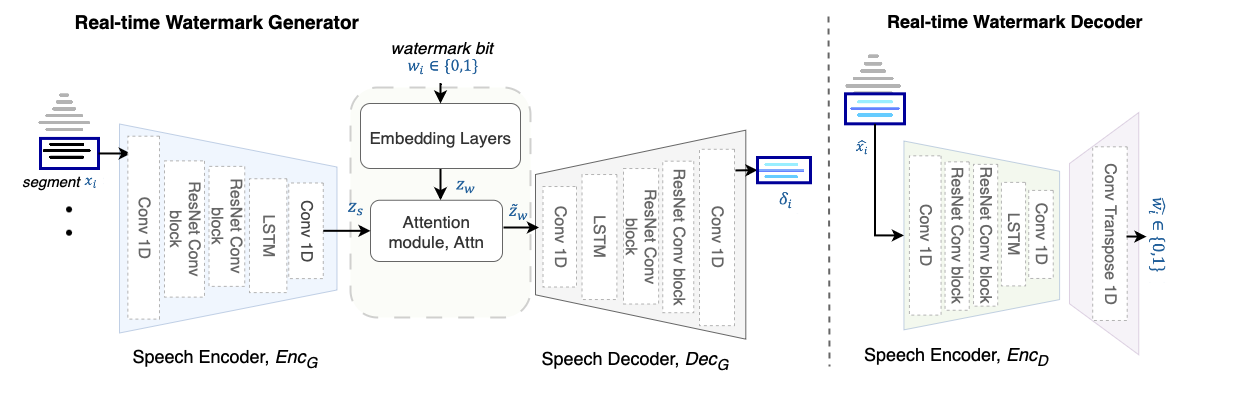}
    \caption{Architecture of the proposed Real-Time Audio Watermarking (RTAW) system.}
    \label{fig:watermark}
\end{figure*}

\heading{Proposed Real-time Audio Watermarking}
In \sys, we introduce a real-time audio watermarking (RTAW) system designed to operate at 8 kHz, with low latency, high robustness, and frame-level granularity. Each protocol message, whether a start beacon, challenge, MAC response, or finish signal, is encoded into a binary sequence and embedded into the speech in real-time using RTAW. 

The proposed RTAW system comprises two core components: \textit{(i)}\textit{ Watermark Generator}, \textit{G} and \textit{(ii)} \textit{Watermark Decoder}, \textit{D} -- designed for real-time frame-by-frame operation to enable low-latency encoding and recovery of protocol messages under realistic telephony conditions. Figure \ref{fig:watermark} provides an overview of the model architecture.

\subsubsection{Real-time Watermark Generator}
\label{sec:system-design-watermark-generator}

The watermark generator, \emph{G}, is designed to embed a hidden message into short, non-overlapping audio segments. Each segment $x_i \in \mathbb{R}^T$ consists of $T=320$ samples, corresponding to 40 ms of audio at an 8 kHz sampling rate. The use of these fixed short segments, or frames, allows for efficient frame-by-frame operation.
We select 40~ms frames to balance watermark robustness with real-time responsiveness. Shorter frames (e.g., 20–30~ms) reduce latency but impair bit detectability; longer frames improve accuracy but introduce unacceptable delays. The 40~ms duration offers a favorable tradeoff, enabling accurate, low-latency, frame-synchronous inference.

\noindent \textbf{Speech Encoder ($Enc_G$):} 
The $x_i$ is first fed into a speech encoder, $Enc_G$, where the signal undergoes multiple downsampling operations to produce a high-dimensional latent frame representation,  $z_s = \mathrm{Enc}_G(x_i)$. 
The encoder architecture includes a 1D causal convolution, followed by a stack of two residual convolutional blocks with depthwise separable convolutions and downsampling layers -- for capturing low-resolution features. For global temporal features, a single LSTM is used, followed by 1D convolution that compresses the output into a compact latent representation $z_s$.

\noindent\textbf{Watermark Integration Module :} 
In parallel, a single-bit message $w_i \in \{0,1\}$ is passed through a learnable embedding layer to produce a richer and dense representation $z_w$. 
An attention mechanism ($Attn$) is used to modulated the watermark representation,  $\tilde{z}_w = \mathrm{Attn}(z_s, z_w)$, conditioned on the speech signal. This ensures that the message is embedded into perceptually and statistically robust regions of the speech.

\noindent \textbf{Speech Decoder ($Dec_G$):} 
The modulated watermark representation $\tilde{z}_w$  is then decoded by $Dec_G$. The decoder $Dec_G$ mirrors the $Enc_G$ architecture with transposed convolutions to upsample the representation and reconstructs a small imperceptible perturbation, $\delta_i = \mathrm{Dec}_G(\tilde{z}_w)$, in the temporal domain. The resultant perturbation $\delta_i$ contains the watermark bit information.

The final watermarked audio segment, $\hat{x}_i \in \mathbb{R}^T$ is obtained by adding the perturbation, $\delta_i$ to the original segment $s_i$. The strength of $\delta_i$ is controlled by a scaling factor, $\alpha \in [0.6,1]$, to balance the robustness of the watermark with its perceptual imperceptibility.
\begin{align}
\hat{x}_i &= x_i + \alpha\delta_i
\end{align}
Both the encoder and decoder architecture are adapted from the EnCodec framework \cite{defossez2022high}. A key design principal is that all the components are strictly causal and frame-synchronous. This ensures there is no dependency on future context, and hence supporting real-time inference.

\subsubsection{Real-time Watermark Decoder}
\label{sec:system-design-watermark-decoder}
The watermark decoder, \emph{D}, is designed to process incoming speech to recover the hidden message sent from the generator. For each watermarked speech segment, $\hat{x}_i $, the \emph{D} predicts the message, $\hat{w}_i$, embedded within the signal. 
The decoder's architecture mirrors the speech encoder of the generator $Enc_G$, with its own unique weights. The input segment $\hat{x}_i $ is first passed through the speech encoder $Enc_D$, followed by a transposed convolution and a linear layer with a softmax function. The \emph{D} outputs the predicted $\hat{w}_i=D(\hat{x}_i)$, $\hat{w}_i\in  \{0,1\}$ for each segment along with detection confidence score indicating the likelihood of the segment containing a valid watermark signal.

\subsection{Data Link Protocol Over Audio}
\label{sec:data_link}
While RTAW enables the embedding of messages into speech, it alone is does not ensure reliable message delivery under realistic telephony conditions. Distortions such as jitter, quantization, echo, filtering, and background noise can corrupt individual bits, leading to frequent decoding failures. 
To address this, and in the absence of standardized protocols for data transmission over watermarked real-time audio, \sys introduces a lightweight data link protocol specifically designed for this setting. We conceptualize the watermarking system as a \emph{noisy serial communication medium}, where each audio carries a single bit that may be flipped during transmission.  To model this behaviour, we consider RTAW as a \emph{Binary Symmetric Channel (BSC)} with crossover probability~$p$, where each bit is independently flipped with probability~$p$. This abstraction allows us to apply well-established coding-theoretic tools to ensure robust message recovery. Built atop this lossy channel, our data link protocol provides essential communication services, including synchronization, error correction, and retransmission.

\begin{figure}[t]
\centering
\begin{tikzpicture}[scale=0.95, every node/.style={scale=0.95}]
  \draw[fill=gray!20, rounded corners=2pt] (0,0) rectangle (2,0.8);
  \node at (1,0.4) {\textbf{\makecell{Sync \\Preamble}}};

  \draw[rounded corners=2pt, thick, draw=black] (2,-0.2) rectangle (7,1.0);
  \node at (4.5,1.15) {\textbf{Payload}};

  \draw[fill=blue!15, rounded corners=2pt] (2,0) rectangle (4.5,0.8);
  \node at (3.25,0.4) {\textbf{Message Bits}};

  \draw[fill=blue!35, rounded corners=2pt] (4.5,0) rectangle (7,0.8);
  \node at (5.75,0.4) {\textbf{ECC Bits}};
  
  \node[anchor=west] at (7.5,0.5) {\shortstack{\textbf{Data Link}\\\textbf{Frame}}};
\end{tikzpicture}
\caption{
Structure of a data link frame transmitted over the watermarking channel.
}
\label{fig:frame_structure}
\end{figure}
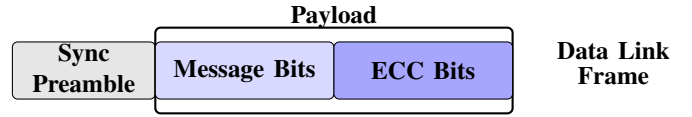

Each transmitted frame (Figure~\ref{fig:frame_structure}) consists of: (1) synchronization preamble, composed of a fixed bit pattern used by the receiver to detect message boundaries via a sliding-window search; and (2) a payload, which contains structured binary data such as start beacons, cryptographic challenges, or responses. To ensure resilience under bit-level corruption, each payload is encoded using a Bose–Chaudhuri–Hocquenghem (BCH) code. Let $M$ denote the original message bits. The encoder produces a longer binary codeword $C$:

\begin{equation}
C = \text{BCH}_{\text{Encode}}(M), \quad C \in \{0,1\}^n
\end{equation}

The codeword $C$ includes both the message and parity bits required for error correction, enabling recovery from up to $t$ bit errors per message. BCH parameters are chosen to balance robustness and bitrate efficiency under real-time telephony constraints.

When a decoding failure occurs, either due to failed synchronization or unrecoverable errors, the sender retransmits only the affected segment, following a protocol similar to stop-and-wait Automatic Repeat reQuest (ARQ) strategies. In this way, the protocol ensures reliability and structure despite the inherent fragility of audio-based bit transmission. To prevent indefinite retry loops, each segment is retransmitted up to a fixed maximum number of attempts (e.g., 3 retries). This bounded retransmission approach ensures reliability and structure, while respecting the latency constraints of real-time audio communication over a fragile watermarking channel.

 By layering synchronization, framing, error correction, and retransmission atop the underlying BSC-modeled channel, the protocol transforms an unreliable low-bitrate stream into a resilient communication channel for real-time telephony authentication.

The data link protocol is content-agnostic. It does not assume any specific semantics for the payload. It supports the transmission of arbitrary binary messages, such as cryptographic tokens, session identifiers, control metadata, or application-layer commands, making it a reusable foundation for secure, low-bitrate communication over audio. This abstraction enables the deployment of diverse higher-layer protocols, including, but not limited to, authentication.

\subsection{Endpoint Authentication Protocol}
\begin{figure}[]
    \centering
    \includegraphics[width=0.95\linewidth]{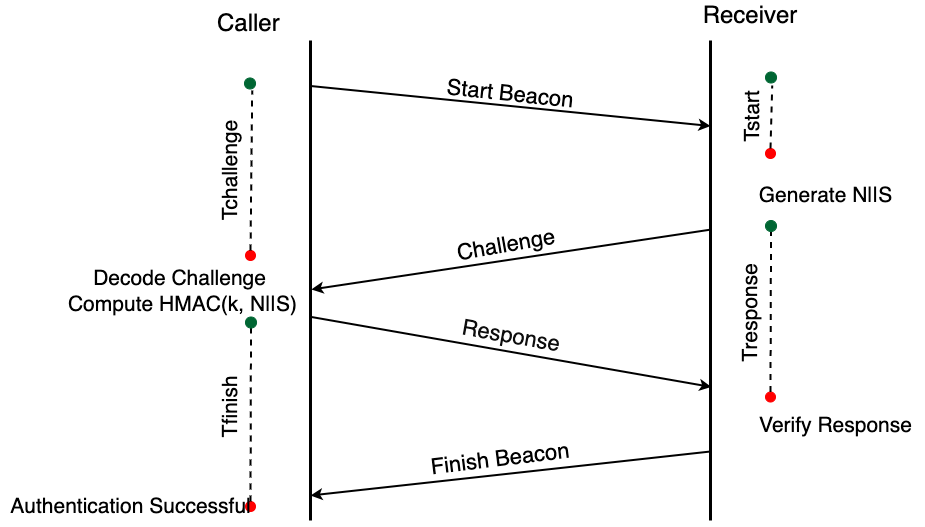}
    \caption{ Overview of phases and message exchange in \sys}
    \label{fig:protocol-diagram}
\end{figure}

We build a challenge–response authentication scheme over its data-over-audio communication channel introduced in Section~\ref{sec:data_link}.  The protocol comprises four phases: Start, Challenge, Response, and Finish, shown in Figure~\ref{fig:protocol-diagram}. Each message is framed with a synchronization preamble and encoded using a BCH code before being embedded into the speech stream via the real-time watermarking layer.

\heading{Start Phase.}
If the caller intends to authenticate the session, it begins by transmitting a predefined start beacon (e.g., \texttt{10110010}) embedded into the outgoing speech using the real-time watermark generator. This beacon is framed using the proposed data link protocol, with a synchronization preamble and BCH-encoded payload to support alignment and error correction.
If the caller does not initiate authentication, no beacon is sent. On the receiver side, the incoming audio is scanned in real time. If a valid beacon is detected within a timeout window $T_{\text{start}}$, the protocol proceeds to the next phase. Otherwise, the receiver concludes that the call is unauthenticated and not protected by \sys. Retransmission of the beacon may be triggered if synchronization fails.

\heading{Challenge phase.}
Once the beacon is detected, the receiver generates a random challenge composed of a 64-bit session-specific nonce $N$ and a 64-bit session identifier $S$. The complete challenge message is formed by concatenating both values:
\begin{equation}
    Challenge=N||S
\end{equation}
If no valid response is received within the timeout $T_{\text{response}}$, the receiver may retransmit the challenge message.

\heading{Response phase}
After sending the start beacon, the caller starts a timer $T_{\text{challenge}}$ and listens for the incoming challenge. If the challenge is received within this window, the caller computes a MAC using a shared secret key $K$ and the received challenge components $\hat{N}$ and $\hat{S}$:
\begin{equation}
\text{MAC} = \text{Truncate}_{128}\big(\text{HMAC}(K, \hat{N} \, || \, \hat{S})\big)
\end{equation}
where the function $\text{HMAC}$, denotes Hash-based Message Authentication Code, which uses a shared secret key $K$ and a hash function, SHA-256 to encrypt the challenge (N$||$S). While $\text{Truncate}_{128}(\cdot)$ truncates the output of the HMAC to the first 128 bits (16 bytes). This reduction provides a shorter message suitable for embedding within a limited audio capacity while retaining sufficient cryptographic strength for authentication purposes. Retransmission is attempted if the response fails to decode.

\heading{Finish phase}
After transmitting the challenge, the receiver starts a timeout $T_{\text{response}}$ to await the caller's response. If a valid response is received within this time frame, the receiver decodes the embedded MAC from the audio stream and recomputes the expected MAC:
\begin{equation}
\text{MAC}_{expected} = \text{Truncate}_{128}\big(\text{HMAC}(K, N \, || \, S)\big)
\end{equation}
If the decoded MAC matches the expected value, authentication is successful, and the receiver transmits a final beacon to confirm completion of the protocol. Otherwise, if the MACs do not match, the receiver assumes that an error may have occurred in either the challenge or response transmission. To recover, it initiates a full retransmission starting from the challenge phase. The caller awaits the finish beacon within a timeout window $T_{\text{finish}}$ to confirm authentication; if no beacon is received, it similarly prepares to retry the exchange.

\section{Experiments}
\subsection{Dataset}
We use the Switchboard-1 Telephone Speech Corpus~\cite{LDC97S62} for training and evaluation of \sys.
The dataset contains approximately 260 hours of natural, spontaneous two-sided telephone conversations recorded between 543 speakers from across the United States. Each conversation was initiated and coordinated by an automated system that paired speakers, suggested discussion topics, and recorded audio on separate channels. The dataset's diversity in speaker demographics and discussion topics makes it particularly well-suited for assessing authentication performance under realistic telephony conditions. 
We adopt a standard 60\%, 20\%, 20\% split for training, validation, and testing. The RTAW model is trained and evaluated on the training and validation set respectively, and the full end-to-end protocol is evaluated on the test set.   

\subsection{Evaluation Metrics}
To comprehensively assess the performance of \sys, we employ a set of evaluation metrics categorized into four distinct groups evaluating: \textit{(i)} Watermark robustness, \textit{(ii)} Data link reliability, \textit{(iii)} Authentication robustness, and \textit{(iv)} Audio quality.

\heading{Audio Watermarking Metrics.} The watermark-related metrics quantify the robustness of the designed watermarking system in recovering the embedded bitstream under transmission and distortions.
\begin{itemize}
\item Bit Accuracy: The fraction of correctly recovered bits across all embedded bitstreams. This metric captures the overall reliability of the watermarking system, even when some bit errors occur.
\item Perfect Recovery Rate: The fraction of runs in which the entire embedded bitstream is recovered with zero bit errors. This stricter metric quantifies how often the watermarking system achieves flawless decoding.
\end{itemize}
\heading{Data Link Reliability Metrics.} These metrics evaluate the reliability of synchronization mechanisms that underlie successful message decoding in \sys.
\begin{itemize}
    \item Sync Detection Accuracy: Fraction of runs in which the synchronization preamble is correctly detected. This metric reflects the protocol's ability to align bitstreams under timing uncertainty and channel noise.
    \item Frame success Rate:  Fraction of transmitted frames that are fully and correctly decoded by the receiver. A frame is considered successful only if the entire message is correctly aligned, decoded, and passes error correction. Partial or corrupted decodings are not counted.
\end{itemize}
\heading{Authentication Robustness Metrics.} The authentication-related metrics quantify the protocol's ability to accurately transmit and recover the message payload under adverse channel conditions.
\begin{itemize}
    \item Beacon Detection Rate: Fraction of runs where the start beacon is detected.
    \item Challenge Success Rate: Fraction of runs where the challenge is correctly decoded.
    \item Response Verification Rate: Fraction of runs where the caller's MAC response is correctly verified.
    \item Finish Beacon Rate: Fraction of runs where the final confirmation is successfully received.
    \item Overall Authentication Success: Fraction of runs where all four phases succeed.
\end{itemize}
\heading{Audio Quality Metrics.} To ensure that watermark embedding does not compromise the perceptual quality of the host signal, we report the following objective measures:
\begin{itemize}
    \item Perceptual Evaluation of Speech Quality (PESQ) \cite{rix2001perceptual}: A standardized metric that quantifies perceived audio quality degradation. Higher PESQ scores indicate minimal distortion introduced by watermarking and subsequent transmission.
    \item Short-Time Objective Intelligibility (STOI) \cite{taal2011algorithm}: An intelligibility-oriented metric that evaluates the impact of the watermarking process on speech comprehension. STOI is particularly sensitive to temporal and spectral distortions and complements PESQ in assessing perceptual integrity.
\end{itemize}

\subsection{Experimental Setup}
\label{sec:experimental-setup}


\heading{RTAW Training Parameters} The proposed RTAW system consists of a watermark generator and decoder, jointly trained by adapting the training strategy of AudioSeal. The model was trained using  Adam optimizer (lr=$5\times10^{-5}$) with cosine annealing and 1000-step warmup. Training runs for 300 epochs with 2000 updates per epoch and a batch size of 16.  Training was performed with a fixed watermark strength of $\alpha = 1.0$,
Details of the model configuration are shown in Table~\ref{tab:watermark-config}.
\begin{table}[ht]
    \centering
    \caption{RTAW model configuration (Generator and Decoder)}
    \label{tab:watermark-config}
    \begin{tabular}{ll}
        \toprule
        \textbf{Parameter} & \textbf{Value} \\
        \midrule
        Input channels & 1 \\
        Latent dimension & 128 \\
        Filters & 32 \\
        Residual layers & 2 \\
        LSTM layers & 1 \\
        Causal & True \\
        Compression ratio & 2 \\
        Ratios & [2, 2] \\
        Kernel sizes & 5 (main), 3 (residual), 5 (final) \\
        Activation & ELU \\ 
        Normalization & WeightNorm \\
        Pad mode & constant \\
        \bottomrule
    \end{tabular}
\end{table}

\heading{Distortion Model.}
To assess real-world robustness, we apply a range of realistic distortions that naturally occur in heterogeneous voice communication networks. These distortions are not adversarial in nature; they arise from codec pipelines, transmission artifacts, and environmental conditions. We do not assume that attackers would deliberately introduce distortion—doing so would reduce watermark integrity and increase the likelihood of fraud detection, which goes against the adversary's goal of stealthy impersonation. The distortion types considered include:

\begin{itemize}
    \item \textbf{White noise:} Simulates random electrical or analog interference during the call.
    \item \textbf{Pink noise:} Represents low-frequency environmental or background noise.
    \item \textbf{Echo:} Emulates acoustic feedback, reverberation or speaker leakage. 
    \item \textbf{Lowpass / Highpass / Bandpass filtering:} Mimics the frequency shaping inherent from telephony codec and communication channel limitation.
    \item \textbf{Ducking:} Emulates dynamic range compression or automatic gain control (AGC), common during multitasking or system notifications.
\end{itemize}

\noindent The parameters used for simulating these distortions are randomly selected from pre-defined ranges. These ranges/values are adapted based on the characteristics of telephony channels, the 8 kHz audio sampling rate, and other relevant empirical studies (see Appendix D.2. in \cite{roman2024proactive}). 

\heading{End-to-end \sys Setup.}
Each message in \sys, start beacon, challenge, response, and finish beacon, is embedded independently and transmitted over a simulated telephony channel. During transmission,  we introduce two classes of distortions to emulate real-world network and acoustic conditions:

\begin{enumerate}
    \item \textbf{Random network delay:} sampled uniformly from 0–120~ms to emulate jitter and buffering.
    \item \textbf{Audio distortion:} applied to a randomly selected portion of the message, with coverage ratios of 20\%, 40\%, 60\%, or 80\%.
\end{enumerate}
We use BCH codes with \(m{=}5, t{=}5\) for start and finish messages, enabling 31-bit codewords that correct up to 5 errors. For challenge and response, we adopt \(m{=}9, t{=}55\), encoding 128 bits into 511-bit codewords with up to 55 error corrections. These parameters were empirically selected to balance robustness and latency---larger \(m\) improves error correction but increases codeword length and decoding time, yielding limited gains in practice.
We execute the full \sys protocol over a representative subset of the Switchboard test set.  For each distortion condition, 10 repetitions are run per audio file. A run is deemed successful if all four protocol phases are received, decoded, and verified correctly within their respective timeout windows.


\section{Evaluation}
\label{sec:results}
\subsection{RTAW }
\label{sec:watermark-eval}

RTAW is evaluated by embedding a fixed 16-bit message into clean speech and recovering it after applying simulated distortions. We vary the watermark strength parameter $\alpha \in \{0.6, 0.8, 1.0\}$ to study the tradeoff between robustness and audio quality. Higher watermark strength $\alpha$ improves detectability but increases the risk of audible artifacts. This tradeoff is especially important in telephony, where both imperceptibility and reliable recovery are essential.

Each 16-bit message is embedded frame-by-frame across 16 consecutive 40 ms segments. To simulate a realistic communication channel, we apply distortions to varying fractions of the signal, affecting 20\%, 40\%, 60\%, and 80\% of the signal. The 20\% level reflects typical degradations observed in telephony, such as transient channel noise, compression artifacts, or jitter~\cite{hu2020evaluating,markopoulou2003assessing}. Higher levels of distortion represent increasingly challenging scenarios, including extreme environments. This graduated setup enables a systematic stress test of RTAW across a controlled distortion spectrum. By progressively increasing the affected signal portion, we characterize not only real-world robustness but also identify thresholds at which the system begins to fail, yielding valuable insights into its resilience and operational limits.

Table~\ref{tab:bit_accuracy_ablation} summarizes the bit accuracy, perfect recovery rate, and audio quality for the two different watermark strengths, i.e., $\alpha = 0.6$ and $\alpha = 1.0$, evaluated under 20\% and 80\% distortion coverage. Results for the intermediate setting $\alpha = 0.8$ and additional distortion rates (40\% and 60\%) are provided in Appendix~\ref{sec:appendix-watermark-results} for completeness.

\heading{Bit accuracy.} 
RTAW achieves high bit accuracy even at the lowest watermark strength. With a low watermark strength of $\alpha = 0.6$, \sys achieves above 91\% bit accuracy for most distortion types and distortion rates. For instance, under 80\% ducking, echo, pink noise, or lowpass filtering, accuracy remains above 91\%, and reaches 92\% in clean audio.
Increasing the watermark strength (e.g., $\alpha = 1.0$) provides marginal improvements, raising accuracy to 94\%, but comes at a perceptual cost. 
Specifically, higher $\alpha$ values introduce stronger perturbations into the signal, which begin to affect perceptual quality metrics. For instance, in clean audio, PESQ drops from 4.44 at $\alpha = 0.6$ to 4.25 at $\alpha = 1.0$, while STOI falls from 0.991 to 0.980. A similar trend is observed under more challenging conditions: with 80\% lowpass filtering, PESQ decreases from 3.98 to 3.84, and STOI from 0.917 to 0.910. While these degradations remain below the threshold of perceptual annoyance, they reflect a measurable reduction in speech naturalness. 

\heading{Perfect Recovery Rate.}
While bit accuracy is consistently high, full-message recovery remains challenging. Even a single bit error can invalidate a challenge or MAC response. Perfect recovery rates vary significantly, ranging from 0.67 on clean audio to below 0.2 for aggressive distortions like white noise or bandpass filtering.
These results show that watermarking alone cannot guarantee cryptographic reliability, motivating the need for a structured data link layer that adds synchronization, redundancy, and error correction to ensure robust message delivery and secure authentication.

We observe a clear tradeoff between watermark strength and perceptual quality. At $\alpha = 0.6$, the system achieves near-transparent quality in clean and lightly distorted speech, with PESQ above 4.3 and STOI above 0.98. Even under challenging conditions like echo and white noise, quality remains high (e.g., STOI $\geq$ 0.88 and PESQ $\geq$ 2.7 for echo at 20\%).
In contrast, increasing $\alpha$ to 1.0 introduces consistent degradation, with PESQ dropping by up to 0.8 points and STOI by 0.02–0.05, particularly in echo and bandpass scenarios.
Overall, $\alpha = 0.6$ offers the best tradeoff—maintaining perceptual quality while ensuring robust message recovery. Stronger perturbations yield only marginal gains and are unnecessary for typical telephony settings. We therefore adopt $\alpha = 0.6$ as the default strength in \sys.

\heading{Latency}
Table~\ref{tab:bit-latency} reports the time required to embed and decode individual bits using our watermark system. All measurements were obtained on a machine configured to simulate mobile hardware constraints, with 4 CPU cores and 4GB of RAM.
Our results show that, on average, the generator embeds a bit in 12.06 ms, with a maximum of 21.52 ms and low variability (standard deviation 1.08 ms), indicating stable and predictable performance.  Decoding is faster, averaging 4.23 ms per bit, with a minimum of 3.02 ms and a maximum of 29.37 ms.  These results confirm that both embedding and decoding are safely within real-time audio processing bounds, and the protocol can operate frame-synchronously without perceptible delay.
\begin{table}[h]
\centering
\caption{Per-bit embedding and decoding latency statistics (in milliseconds)}
\label{tab:bit-latency}
\begin{tabular}{lcccc}
\toprule
\textbf{Operation} & \textbf{Avg (ms)} & \textbf{Max (ms)} & \textbf{Min (ms)} & \textbf{Std (ms)} \\
\midrule
Embedding & 12.055 &  21.521 &  9.661 & 1.076  \\
Decoding& 4.225  & 29.366  &  3.0224 & 0.57   \\
\bottomrule
\end{tabular}
\end{table}

\heading{Takeaways}
This evaluation reveals three key insights. First, RTAW achieves strong bit-level robustness across diverse distortions, even at low strength. Second, perfect message recovery is far more fragile; rare bit errors are sufficient to break authentication. Third, these limitations highlight that watermarking alone is insufficient for secure communication.
\sys addresses this by layering synchronization, redundancy, and BCH error correction atop the watermarking primitive, forming a structured data link protocol that enables reliable, delay-tolerant message exchange.

\begin{table*}[t]
\centering
\caption{Bit accuracy, perfect recovery, and audio quality (PESQ/STOI) of RTAW }
\label{tab:bit_accuracy_ablation}
\begin{tabular}{clrrrrrrrr}
\toprule
\textbf{Distortion} & \textbf{\makecell{Rate  (\%)}} &
\multicolumn{2}{c}{\textbf{ \makecell{Bit Accuracy}}} &
\multicolumn{2}{c}{\textbf{\makecell{Perfect Recovery}}} &
\multicolumn{2}{c}{\textbf{Q ($\alpha{=}0.6$)}} &
\multicolumn{2}{c}{\textbf{Q ($\alpha{=}1.0$)}} \\
\cmidrule(lr){3-4} \cmidrule(lr){5-6} \cmidrule(lr){7-8} \cmidrule(lr){9-10}
& & $\alpha{=}0.6$ & $\alpha{=}1.0$ & $\alpha{=}0.6$ & $\alpha{=}1.0$ & PESQ & STOI & PESQ & STOI \\
\midrule
Clean             & -- & 0.920 & 0.981 & 0.670 & 0.850 & 4.438 & 0.991 & 4.245 & 0.980 \\
Bandpass          & 20 & 0.898 & 0.968 & 0.275 & 0.625 & 3.985 & 0.923 & 3.702 & 0.913 \\
                  & 80 & 0.835 & 0.930 & 0.200 & 0.325 & 3.520 & 0.793 & 3.407 & 0.775 \\
\makecell{DuckAudio} & 20 & 0.921 & 0.982 & 0.425 & 0.700 & 4.387 & 0.984 & 4.205 & 0.971 \\
                       & 80 & 0.917 & 0.981 & 0.425 & 0.725 & 4.347 & 0.970 & 4.179 & 0.965 \\
Echo              & 20 & 0.917 & 0.980 & 0.400 & 0.650 & 2.782 & 0.897 & 2.741 & 0.888 \\
                  & 80 & 0.912 & 0.971 & 0.325 & 0.675 & 1.734 & 0.684 & 1.679 & 0.668 \\
Highpass          & 20 & 0.917 & 0.981 & 0.425 & 0.750 & 3.633 & 0.917 & 3.527 & 0.910 \\
                  & 80 & 0.906 & 0.978 & 0.400 & 0.625 & 3.291 & 0.771 & 3.206 & 0.763 \\
Lowpass           & 20 & 0.914 & 0.975 & 0.375 & 0.675 & 4.091 & 0.967 & 3.920 & 0.956 \\
                  & 80 & 0.879 & 0.961 & 0.250 & 0.550 & 3.978 & 0.917 & 3.843 & 0.910 \\
\makecell{PinkNoise} & 20 & 0.924 & 0.982 & 0.375 & 0.700 & 4.269 & 0.978 & 4.151 & 0.967 \\
                       & 80 & 0.924 & 0.980 & 0.375 & 0.675 & 4.103 & 0.960 & 4.025 & 0.951 \\
\makecell{WhiteNoise} & 20 & 0.881 & 0.958 & 0.100 & 0.325 & 3.310 & 0.926 & 3.292 & 0.918 \\
                        & 80 & 0.762 & 0.889 & 0.000 & 0.025 & 2.768 & 0.881 & 2.773 & 0.880 \\
\bottomrule
\end{tabular}
\end{table*}

\subsection{Data Link Protocol Over Audio }

The synchronization pattern used in the proposed datalink protocol is critical for aligning message decoding in noisy, delayed channels. Each protocol message is preceded by a fixed-length preamble constructed by repeating a short binary base pattern. While the use of preambles is introduced in our setup (Section~\ref{sec:experimental-setup}), we focus here on how their structure affects performance and overhead. 
 We evaluate a diverse set of 15 sync candidates, listed below, which vary in both base pattern and number of repetitions:
\begin{itemize}
    \item \textbf{Short repeated patterns (3 bits)}:
    \texttt{[0,0,0]}, \texttt{[1,1,1]}, \texttt{[1,0,1]}, \texttt{[0,1,0]} with 3× and 5× repetition.
    \item \textbf{Complex patterns (5 bits)}:
    \texttt{[1,1,0,1,0]}, \texttt{[1,0,1,0,1]}, \texttt{[0,1,0,1,0]}, \texttt{[1,1,0,0,1]}, \texttt{[1,0,0,1,1]}, \texttt{[0,1,1,0,0]}, \texttt{[1,0,1,1,0]},  all with 3× repetition.
\end{itemize}

Table~\ref{tab:sync-all} summarizes the sync detection accuracy for all 15 evaluated patterns, showing performance on clean audio and the average across eight distortion types. Each pattern was tested across multiple audio files with 10 randomized trials per condition, ensuring statistical reliability. Full per-distortion breakdowns appear in Appendix~\ref{sec:appendix-sync-patterns}.

We observe that both pattern entropy and repetition length play a key role in synchronization robustness. The highest-performing pattern, \texttt{[0,1,0,1,0] × 3}, achieved 100\% accuracy in clean audio and 99\% on average under distortion. Similarly, \texttt{[1,0,1] × 5} reached 98\% under distortion, indicating that high-entropy patterns with rich transition structure are more robust across noise, filtering, echo, and dynamic range effects.

In contrast, low-entropy patterns such as \texttt{[0,0,0] × 3} and \texttt{[1,1,1] × 3} performed poorly, with average sync accuracy as low as 75\% and 65\%, respectively. These uniform sequences are more likely to be confused with silence, periodic speech energy, or consistent background noise.
Longer repetitions generally improve sync robustness. For example, increasing \texttt{[0,0,0]} from 3× to 5× improves detection from 75\% to 80\%.

\begin{table}[h]
\centering
\caption{Sync detection accuracy for all evaluated patterns under clean and average distortion conditions.}
\label{tab:sync-all}
\scriptsize
\begin{tabular}{lcc}
\toprule
\textbf{Pattern} & \textbf{Clean} & \textbf{Dist} \\
\midrule
0 1 0 1 0 x3     & 1.00 & 0.99 \\
1 0 1 x5         & 1.00 & 0.98 \\
0 1 0 x3         & 0.95 & 0.95 \\
0 1 0 x5         & 0.95 & 0.95 \\
0 1 1 0 0 x3     & 0.95 & 0.95 \\
1 0 1 0 1 x3     & 0.95 & 0.94 \\
1 0 1 x3         & 0.95 & 0.94 \\
1 1 0 0 1 x3     & 0.95 & 0.92 \\
1 1 0 1 0 x3     & 0.95 & 0.96 \\
1 0 1 1 0 x3     & 0.90 & 0.91 \\
1 0 0 1 1 x3     & 0.85 & 0.86 \\
0 0 0 x5         & 0.80 & 0.79 \\
0 0 0 x3         & 0.75 & 0.75 \\
1 1 1 x3         & 0.65 & 0.65 \\
1 1 1 x5         & 0.65 & 0.66 \\
\bottomrule
\end{tabular}
\end{table}

Following the synchronization pattern analysis, we evaluate the impact of individual data link components, synchronization, and error correction, using an ablation study focused on 32 bits message. This analysis isolates the role of each mechanism on clean audio, where distortions are absent and decoding errors are due only to alignment drift or random bit flips.

We compare three configurations: (i) No Sync, (ii) Sync Only, and (iii) Full Protocol with both synchronization and BCH-based error correction. In the absence of synchronization, decoding fails entirely (0\% frame success rate). Adding sync preambles enables partial recovery (68.5\% frame success rate), while the full protocol achieves 98\% frame success rate on the first attempt. These results underscore the necessity of both temporal alignment and forward error correction, even in distortion-free conditions.

\heading{Takeaways.}
Our evaluation yields two key insights for designing reliable data link protocol over the watermark channel. First, synchronization pattern structure is critical: high-entropy, asymmetric patterns (e.g., \texttt{[0,1,0,1,0]}) achieve high detection accuracy under distortion, while uniform patterns are prone to false positives. \sys adopts a 15-bit sync preamble that balances robustness and latency.

Second, the ablation study shows that both synchronization and error correction are necessary—even in clean audio. Without sync, decoding fails entirely; without error correction, residual bit errors limit reliability. The full protocol with BCH decoding achieves 98\% success on first attempt. Retransmission adds resilience with minimal delay.

These results confirm that sync, error correction, and bounded retransmission are essential for robust data delivery over a noisy audio channel.

\subsection{Endpoint Authentication Protocol}

\heading{\sys on single transmission.} 
We start by evaluating the robustness of \sys under a single transmission attempt, i.e., without retransmission or fallback. This setup reflects a conservative design assumption in which retrying is either not permitted (due to latency or resource constraints) or not yet triggered. It allows us to isolate the protocol's intrinsic resilience to distortion and quantify first-attempt reliability. These results serve as a baseline for understanding where failures originate and motivate the need for retransmission mechanisms described next.
\begin{figure*}[tb]
    \centering

    \begin{subfigure}[t]{0.33\textwidth}
        \includegraphics[width=\linewidth]{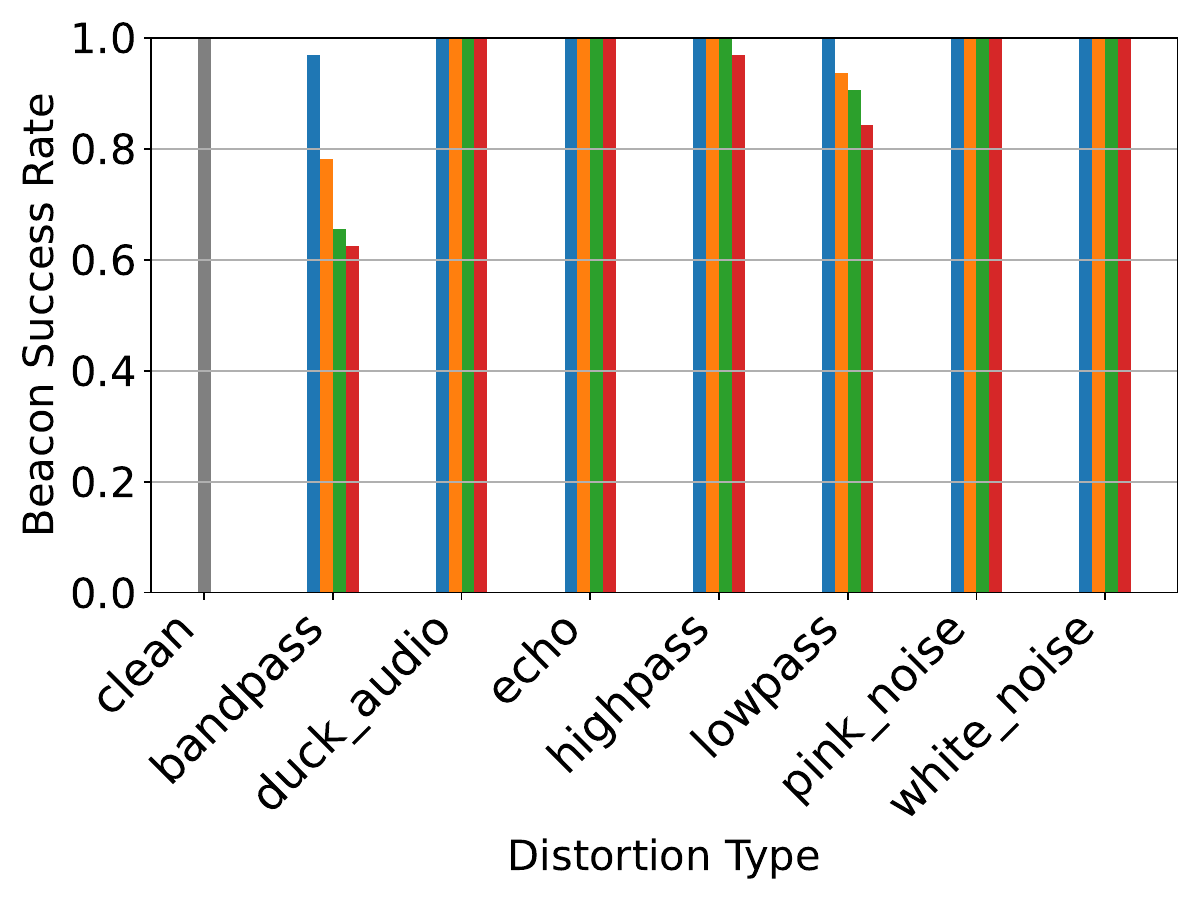}
        \caption{Beacon Success}
    \end{subfigure}
    \begin{subfigure}[t]{0.322\textwidth}
        \includegraphics[width=\linewidth]{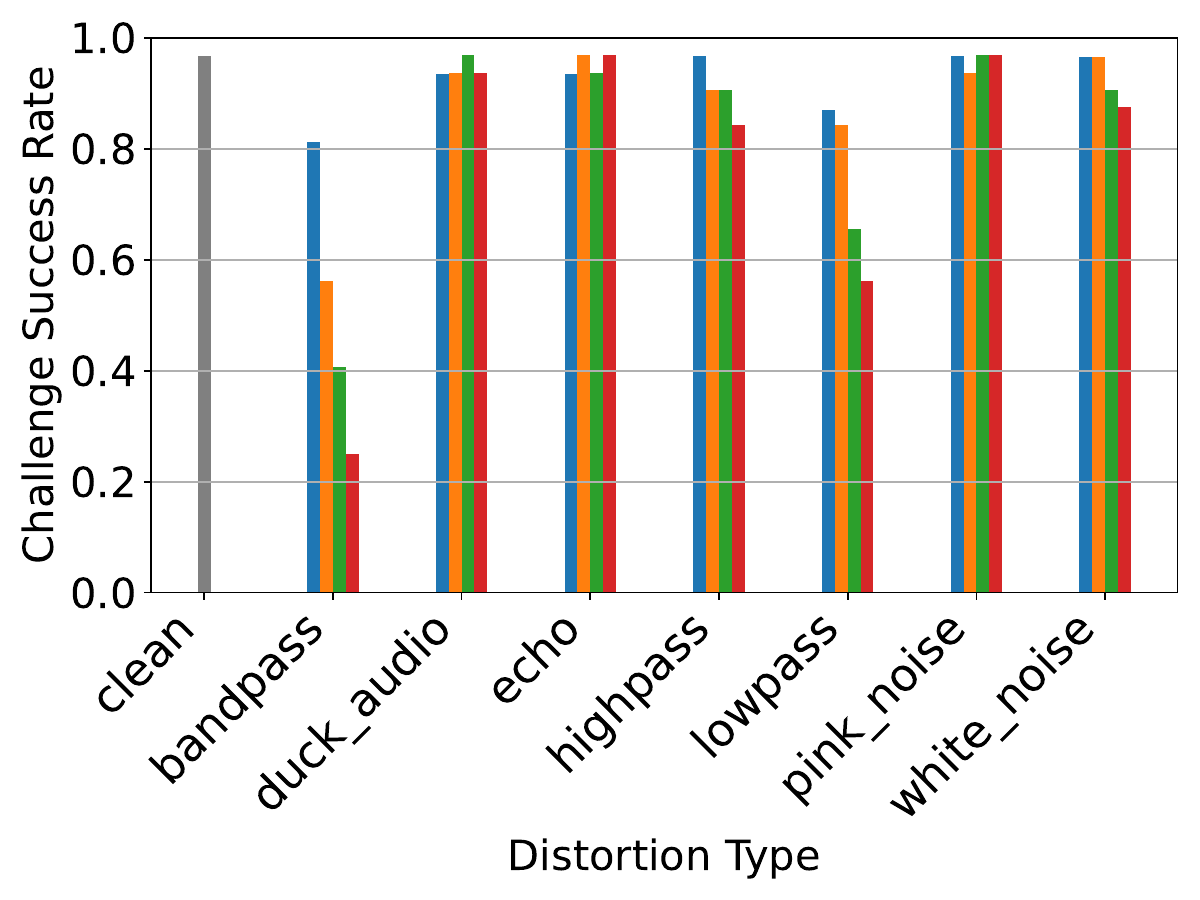}
        \caption{Challenge Success}
    \end{subfigure}
    \begin{subfigure}[t]{0.33\textwidth}
        \includegraphics[width=\linewidth]{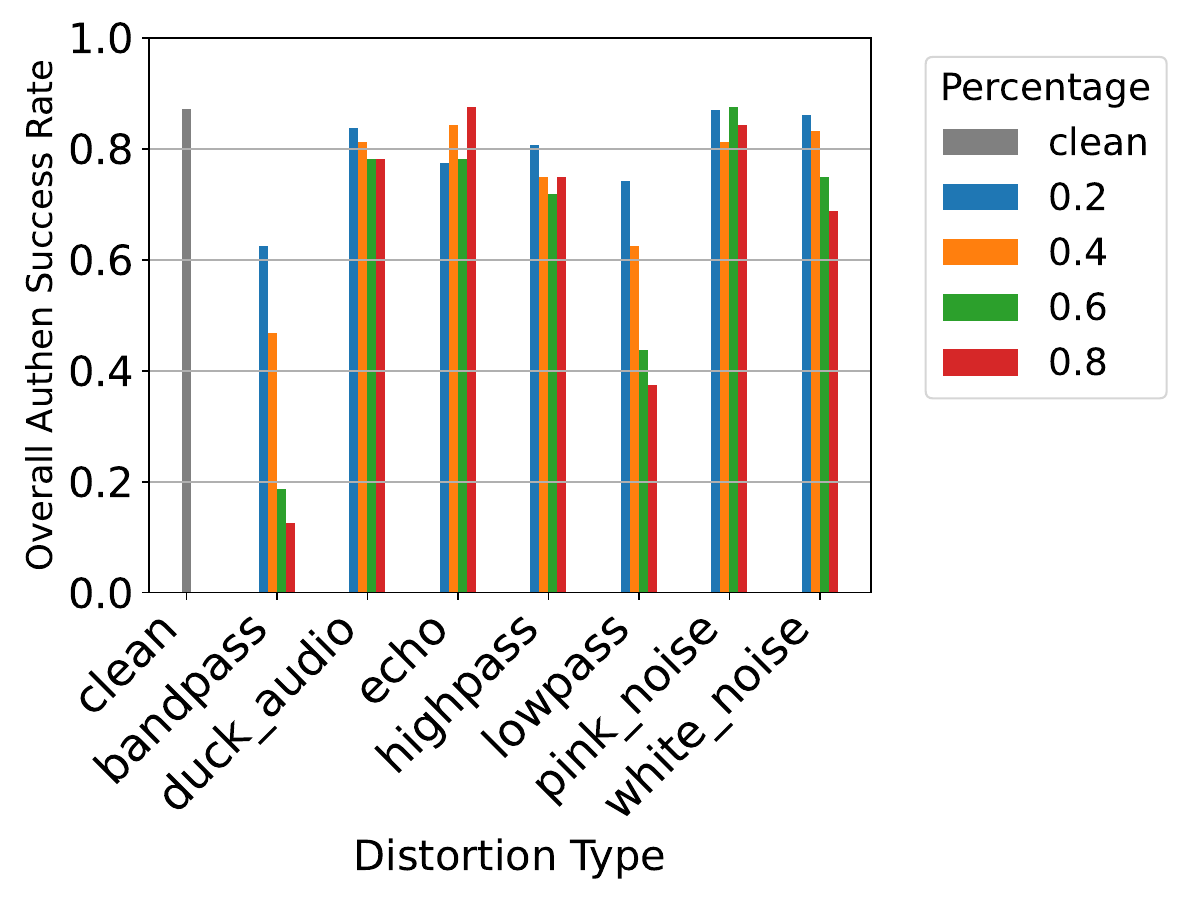}
        \caption{Overall authetictation Success}
    \end{subfigure}

    \caption{Success rates for key protocol stages and overall authentication under varying distortions. Additional per-stage plots appear in Appendix~\ref{sec:appendix-protocol-plots}.}
    \label{fig:protocol-success-main}
\end{figure*}
Figure~\ref{fig:protocol-success-main} shows the success rates of key protocol stages—\emph{beacon}, \emph{challenge}, and the overall end-to-end pipeline—under varying distortion types and corruption intensities. Under clean conditions, \sys achieves an 86\% authentication success rate, compared to 67\% with a single watermark message. This improvement reflects the need of the data link protocol, which integrates synchronization patterns and BCH error correction to create a delay-tolerant and error-resilient data link layer atop the real-time watermarking system.

The protocol operates sequentially: each stage executes only if the previous one succeeds. If the beacon is not detected, no challenge is sent; if the challenge fails, the response is skipped; and if the response is invalid, the finish beacon is suppressed. As a result, each stage's success rate reflects both its own robustness and its dependency on earlier stages.

Beacon detection is consistently the most reliable, achieving over 90\% success under moderate distortion ($\leq$60\%) due to its short payload and strong synchronization cues. Challenge and response stages are more vulnerable to distortion, particularly under white noise or echo. These effects reduce their success rates to 70–75\% under high distortion. Finish beacon performance closely mirrors the response phase, since it is only transmitted after successful MAC verification.

Distortion-specific trends reveal distinct failure modes. Pink noise, ducking, and highpass filtering have minimal impact, with success rates above 80\% across stages. Lowpass filtering is more disruptive, particularly for beacon and challenge phases, as it attenuates high-frequency components needed for sync detection. White noise and echo degrade both synchronization and payload decoding by reducing spectral contrast. Bandpass filtering results in the most severe degradation: at just 20\% coverage, success rates fall below 70\% across all stages due to suppression of midband frequencies essential for watermark energy. However, such extreme filtering is uncommon in real-world telephony, where codecs (e.g., AMR-NB~\cite{3gpp_amr}, EVS~\cite{3gpp_evs}) preserve midband content for speech intelligibility.

Additional per-stage plots for response and finish are provided in Appendix~\ref{sec:appendix-protocol-plots}.

To further understand robustness without retries, we analyze first-attempt failures. The challenge stage accounts for 39.0\% of failures, followed by response (26.8\%) and finish (16.2\%). Beacon failures are rare (9.0\%), confirming that synchronization is generally reliable, and that decoding long payloads remains the primary vulnerability, especially under bursty or frequency-selective distortion.

We analyze failures at the first attempt (i.e., $\alpha = 0.6$) and find that many occur in spectrally sparse audio, such as silence, long low-pitch vowels, or tonal background noise (Appendix~\ref{sec:appendix-failures}). These types of audio lack the high-frequency detail needed for strong watermark embedding, making it harder for the decoder to extract the message correctly when the watermark is weak.

\heading{\sys with Retransmission}

To address transient failures caused by distortion or misalignment, \sys implements a bounded retransmission strategy inspired by ARQ protocols. Rather than repeating the full authentication sequence, it selectively retries only the failed segment, up to a maximum of three attempts, preserving responsiveness for real-time scenarios.

Retransmission is evaluated under a fixed distortion intensity of 20\%, applied uniformly across all distortion types. This level reflects typical degradation observed in telephony environments and enables a consistent comparison across retry strategies.

We compare two approaches: (1) \emph{constant-strength retransmission}, where the watermark strength $\alpha$ remains fixed across attempts, and (2) \emph{adaptive-strength retransmission}, where $\alpha$ increases incrementally after each failure, up to 1.0. In both cases, “1 attempt” indicates first-try success, while “2” and “3” correspond to one and two retries.

\begin{figure*}[t]
  \centering
  \begin{subfigure}[b]{0.48\linewidth}
    \centering
    \includegraphics[width=\linewidth]{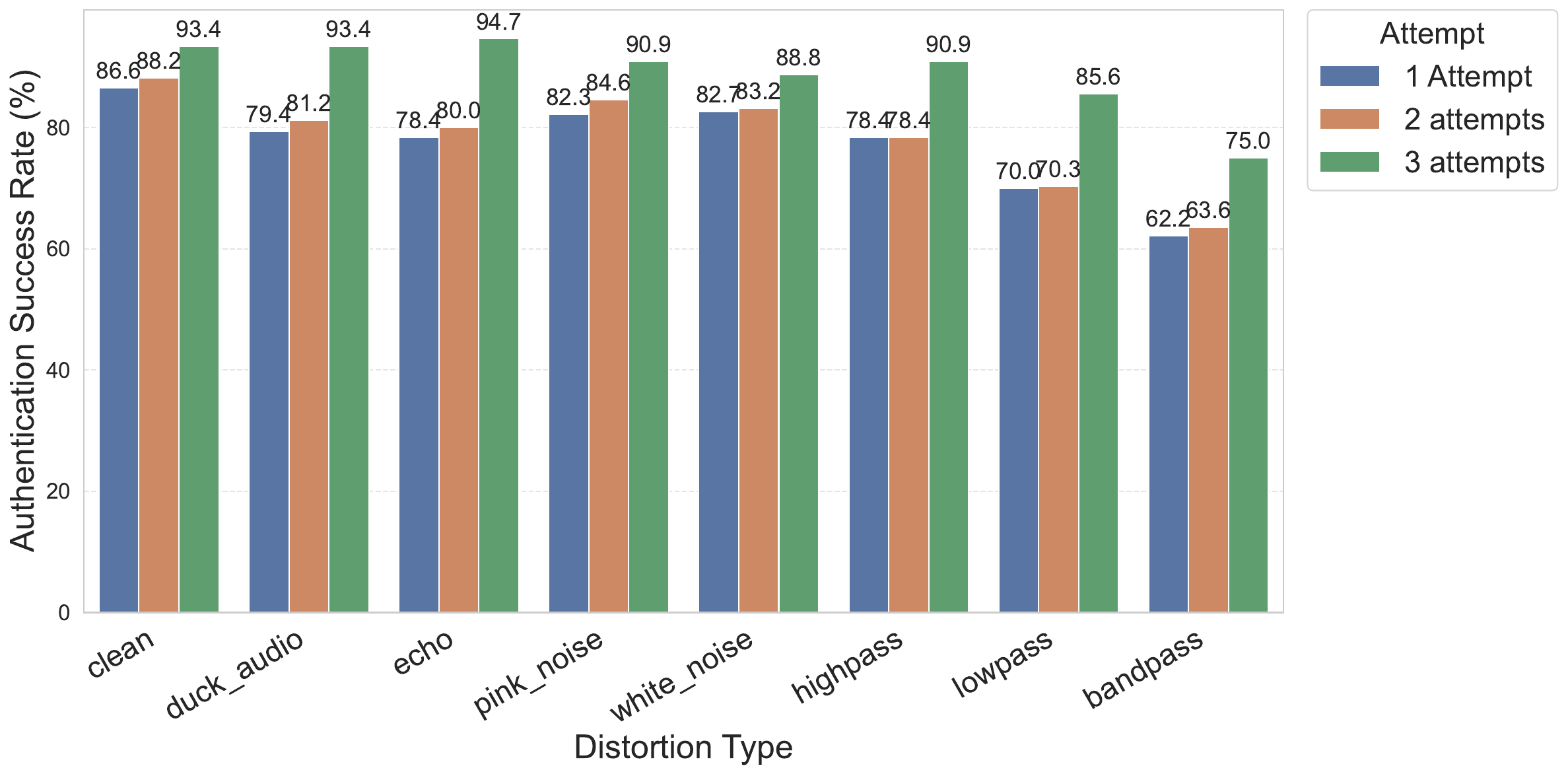}
    \caption{Constant-strength ($\alpha$) strategy. Retransmissions repeat with the same $\alpha$ value.}
    \label{fig:retry-constant}
  \end{subfigure}
  \begin{subfigure}[b]{0.48\linewidth}
    \centering
    \includegraphics[width=\linewidth]{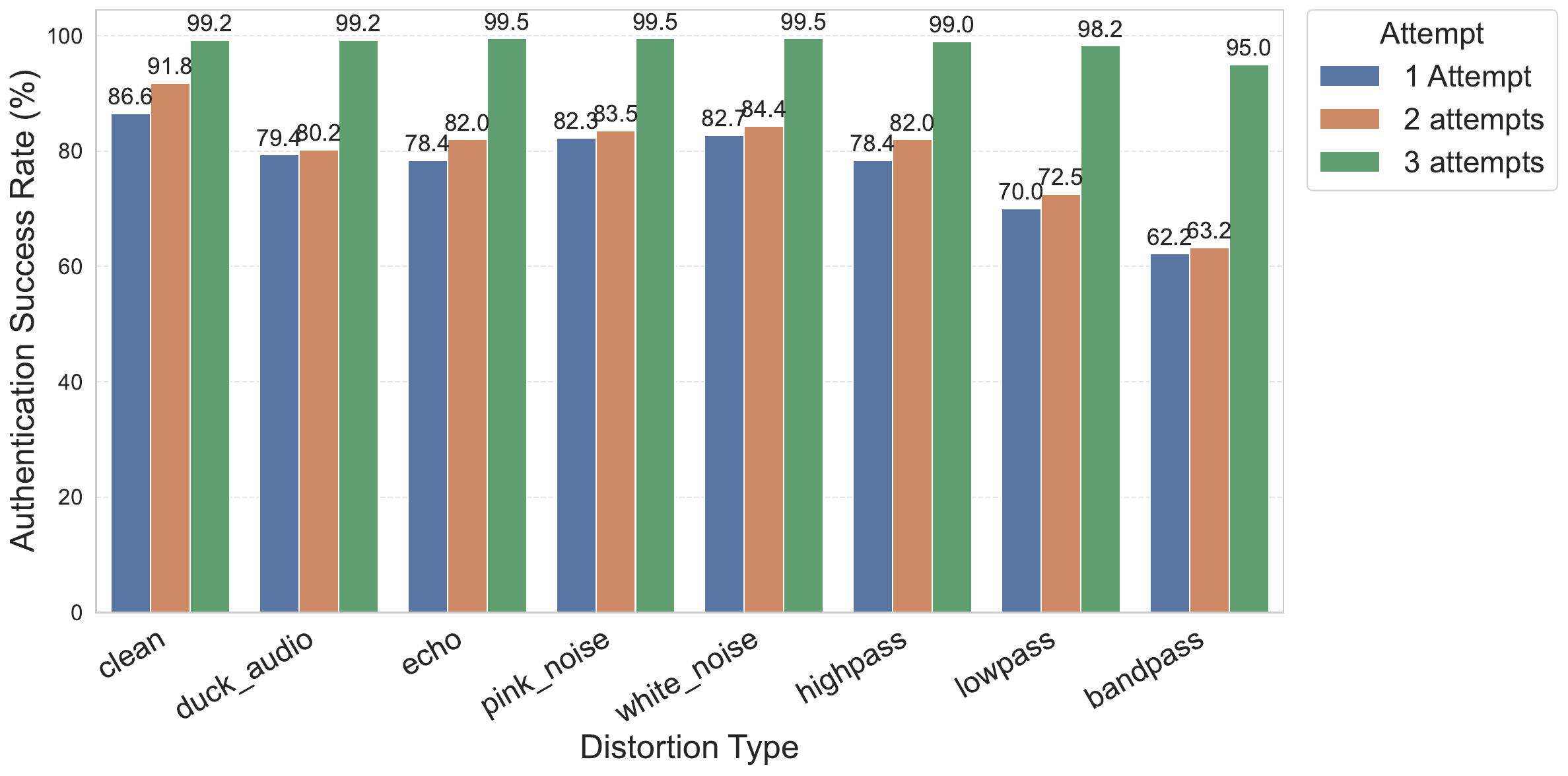}
    \caption{Adaptive-strength strategy. Retransmissions use increasing $\alpha$ values.}
    \label{fig:retry-variable}
  \end{subfigure}
  \caption{Authentication success rates across retries and distortions under different watermark strength strategies.}
  \label{fig:retry-comparison}
\end{figure*}

Figure~\ref{fig:retry-constant} shows results for the constant-strength strategy. Even without increasing $\alpha$, retransmissions significantly improve reliability—for example, boosting clean audio success from 86.4\% to 93.4\%. Under more challenging distortions like lowpass and bandpass filtering, success improves from 70.0\% and 62.2\% to 85.6\% and 75.0\%, respectively.

Figure~\ref{fig:retry-variable} presents results for adaptive-strength retransmission, where $\alpha$ is incrementally increased by 0.1 after each failure. This progressive fallback strategy significantly improves robustness, especially under heavy degradation. For clean audio, the first attempt already achieves over 86\% success; a single retry boosts success to over 91\%. In more difficult cases, such as echo and highpass filtering, success rates exceed 99\% after two retries. Under the most severe conditions—including bandpass and lowpass filtering—success improves from 62.2\% and 70.0\%, respectively, to 95.0\% and 98.2\% after two retries.

Beyond overall reliability gains, adaptive retransmission improves robustness by gradually increasing watermark strength after each failure. While some recoveries occur at $\alpha = 0.7$ during the first retry, a substantial portion of authentications—especially under severe conditions like bandpass and lowpass filtering—are only recovered at $\alpha = 0.8$ in the second retry. This confirms that while moderate strength increases help, escalating to higher $\alpha$ values is often necessary to overcome persistent distortion and ensure reliable bit recovery.

To better understand how watermark strength ($\alpha$) affects detectability, we visualize the spectrograms of embedded audio at different $\alpha$ levels (Appendix~\ref{sec:appendix-bit-corrections}). At $\alpha = 0.6$, the watermark signal is weak and spread out, making it harder for the decoder to detect and leading to more bit errors. At $\alpha = 0.8$, the watermark becomes stronger and more concentrated, especially in mid-frequency regions. These clearer patterns match the decoder's learned filters more effectively, improving bit recovery.

\heading{Retransmission vs Time Trade-off}

While retransmission improves success rates, it may increase overall latency. Table~\ref{tab:auth-time-retries} shows the total time required to successfully authenticate as a function of retry count. The median latency remains bounded between 60–65 seconds, thanks to our partial retransmission design: only the failed segment (e.g., challenge–response or finish) is retransmitted rather than restarting the entire protocol. A small number of outlier runs exceed 80 seconds; these correspond to cases where early failures (e.g., repeated beacon loss) force full re-execution of the protocol sequence. Overall, the results confirm that retransmission adds robustness with minimal overhead.

\begin{table}[h]
  \centering
  \caption{Authentication Time Statistics by Number of Retries}
  \label{tab:auth-time-retries}
  \begin{tabular}{lccc}
    \toprule
    \textbf{Retries} & \textbf{Mean (s)} &  \textbf{Min (s)} & \textbf{Max (s)} \\
    \midrule
    1 attempt & 54.8 & 54.8& 54.8  \\
    2 attempts & 61.52 & 56.64 & 80.36  \\
    3 attempts & 62.54 & 58.48 & 105.92  \\
    \bottomrule
  \end{tabular}
\end{table}
\heading{Takeaways.} The results demonstrate that \sys's retransmission mechanism significantly enhances reliability across all distortion conditions while maintaining low perceptual and temporal overhead. First, retrying with the same watermark strength already offers strong benefits, recovering most failures from transient decoder uncertainty. Second, adaptive-strength retransmission further increases robustness, particularly under severe degradations, while using higher watermark strength only when necessary. Third, time overhead remains acceptable, with most authentications completing within a minute—even when retries occur. These findings validate \sys's practical deployability for secure caller authentication in real-time telephony environments.
\subsection{Security and Overhead Analysis of \sys }
\label{sec:security-analysis}
\heading{Security analysis}
We evaluate the security guarantees and limitations of \sys in the context of the threat model defined in Section~\ref{sec:threat-model}, focusing on its ability to resist impersonation, spoofing, and active manipulation by adversaries controlling the call's origin.

\textit{Resistance to Impersonation and Caller ID Spoofing.}
\sys protects against spoofing by binding authentication to a challenge–response protocol computed over a shared secret. Caller ID metadata is not trusted; instead, identity is tied to the ability to produce a valid cryptographic response. The attacker cannot impersonate a trusted entity unless they possess the secret key, regardless of voice similarity, injected audio, or spoofed SIP/SS7 headers.

\textit{Replay and injection attacks.}
To prevent replay, \sys uses a 64-bit challenge nonce embedded during each session. The MAC is computed over this fresh challenge, ensuring that responses from previous calls are invalid. Even if an adversary records a watermarked message from a prior session, they cannot reuse it successfully. Similarly, pre-recorded or synthesized audio (e.g., deepfakes) without access to the session's live challenge and key cannot pass verification. Hence,  \sys achieves replay resistance by design.

\textit{Adversarial Watermarking and Message Forgery.}
We assume that \sys is public.  An attacker can run their own instance of \sys and embed a custom bitstream into audio. However, without access to the shared secret key, the adversary cannot compute a valid MAC, and the message will fail verification. 
Authentication in \sys relies on embedding an HMAC-SHA256 over a session-specific challenge. As long as the key remains secret, the attacker has no practical way to forge a valid response. This holds even if the adversary knows the full protocol, has access to previous messages, and controls the audio pipeline. Security relies on the widely accepted assumption that HMAC cannot be forged without the key.

\heading{Overhead analysis }
We evaluate the overheads introduced by \sys across three key axes: communication bandwidth, computational cost, and latency.

\textit{Communication Overhead.} \sys introduces minimal overhead by embedding one bit per 40 ms audio frame, resulting in a watermarking capacity of 25 bps. This is significantly below the estimated perceptual capacity of narrowband speech and does not impact the channel’s primary purpose of speech delivery. Furthermore, since watermark bits are embedded directly into the audio stream, the system adds no extra communication traffic at the network layer.

\textit{Computational Overhead.} Watermark embedding and decoding are performed using lightweight models. On a 4-core CPU, each watermark bit is embedded in under 15 ms and decoded in under 5 ms, supporting real-time inference on commodity hardware. Cryptographic operations (e.g., HMAC-SHA256 for MAC generation and verification) are computationally negligible, completing in 1 ms. 
\textit{Latency.} Protocol execution incurs minimal added latency. We conservatively add fixed transmission delays to support sync detection and frame alignment. These delays are absorbed within the normal buffering range of typical telephony pipelines, and authentication completes within an average of 63 s of natural conversation without interrupting or delaying speech flow.

\subsection{Limitations and Extensions}
While \sys demonstrates strong robustness and low-latency performance for narrowband telephony, it currently assumes deterministic cryptographic operations and fixed-bandwidth audio. First, our MAC-based authentication scheme relies on deterministic keyed hashes, which are vulnerable to replay under certain active adversaries if session uniqueness is not enforced. Extending \sys to support probabilistic cryptographic primitives would improve forward security and replay resistance. Second, while \sys assumes a binary symmetric channel model, it does not explicitly address burst errors. Future work could incorporate a more expressive channel model, such as the Gilbert–Elliott model, and adopt error correction strategies suited to bursty conditions.

\section{Conclusion}
\sys introduces a real-time caller authentication system for live telephony, built around three integrated components. First, it employs a real-time audio watermarking model that embeds authentication bits directly into narrowband speech with low latency and high audio transparency. Second, it treats the watermarking pipeline as a noisy communication channel and implements a data link protocol that provides synchronization, error correction, and bounded retransmission for robust bit delivery. Third, it layers a lightweight authentication protocol atop this channel, using symmetric-key challenge–response exchanges to verify caller identity. The protocol relies on pairwise secrets shared offline between trusted parties (e.g., individuals and their banks or agencies), and uses a Hash-based Message Authentication Code (HMAC) to prove that the caller possesses the correct secret key—ensuring cryptographic binding between the audio stream and the claimed identity.

Our evaluations demonstrate that \sys achieves over 99.2\% end-to-end authentication success on clean audio and maintains strong robustness across diverse degradations, while preserving imperceptible audio quality. Unlike prior watermarking systems that operate offline or require high-fidelity audio, \sys runs on narrowband (8\,kHz) speech in real-time and completes authentication in an average of 63 seconds per call. Additionally, it ensures a 0\% false acceptance rate against adversaries without the shared key.

Beyond its standalone utility, \sys lays the foundation for secure telephony overlays in untrusted environment. Future work includes integrating probabilistic cryptographic primitives to improve replay resistance and forward security, and extending the channel model to handle burst errors.
\textbf{Availability.} To support reproducibility and future research, we will release the full implementation of \sys—including the watermarking models, data link protocol, and authentication pipeline.






\bibliographystyle{IEEEtran}
\bibliography{refs}
%



\appendices
\section{Extended Watermark Evaluation Results}
\label{sec:appendix-watermark-results}

To complement the main evaluation in Section~\ref{sec:watermark-eval}, Table~\ref{tab:bit_accuracy_results_full} presents the full set of results across all watermark strengths $\alpha \in \{0.6, 0.8, 1.0\}$ and all distortion coverage rates {20\%, 40\%, 60\%, 80\%}. These results are included for completeness and to enable finer-grained analysis of robustness trends across different operating points.

The intermediate watermark strength $\alpha = 0.8$ consistently performs between the low and high extremes: it offers modest improvements in bit accuracy and perfect recovery compared to $\alpha = 0.6$, but with smaller perceptual degradation than $\alpha = 1.0$. Similarly, distortion rates of 40\% and 60\% bridge the gap between typical telephony noise levels (20\%) and stress-test conditions (80\%), revealing gradual declines in performance as distortions increase.

While these intermediate values provide useful diagnostic insight, they are excluded from the main discussion for clarity and space. 
\begin{table*}[t]
\centering
\caption{Bit accuracy of the watermark system for each distortion type across watermark strengths $\alpha$ and distortion coverage rates.}
\label{tab:bit_accuracy_results_full}
\begin{tabular}{|ll|ccc|lll|llllll|}
\hline
\multicolumn{2}{|c|}{\textbf{Distortion}}                                                            & \multicolumn{3}{c|}{\textbf{Bit Accuracy}}                                                    & \multicolumn{3}{l|}{\textbf{Perfect Recovery rate}}                                                  & \multicolumn{6}{c|}{\textbf{Audio Quality}}                                                           \\ \hline
\multicolumn{1}{|c}{\multirow{2}{*}{\textbf{Distortion Type}}} & \multirow{2}{*}{\textbf{Rate (\%)}} & \multirow{2}{*}{$\alpha=0.6$} & \multirow{2}{*}{$\alpha=0.8$} & \multirow{2}{*}{$\alpha=1.0$} & \multirow{2}{*}{$\alpha=0.6$} & \multirow{2}{*}{$\alpha=0.8$} & \multirow{2}{*}{$\alpha=1$} & \multicolumn{2}{c}{$\alpha=0.6$} & \multicolumn{2}{l}{$\alpha=0.8$} & \multicolumn{2}{l|}{$\alpha=1$} \\ \cline{9-14} 
\multicolumn{1}{|c}{}                                          &                                     &                               &                               &                               &                               &                               &                             & PESQ            & STOI           & PESQ            & STOI           & PESQ           & STOI           \\ \hline
Clean                                                          & -                                   & 0.920                         & 0.962                         & 0.981                         & \multicolumn{1}{c}{0.67}       & \multicolumn{1}{c}{0.825}       & \multicolumn{1}{c|}{0.850}    & 4.4380          & 0.991          & 4.353           & 0.986          & 4.245          & 0.980          \\ \hline
Bandpass                                                       & 20                                  & 0.898                         & 0.944                         & 0.968                         & \multicolumn{1}{c}{0.275}     & \multicolumn{1}{c}{0.375}     & \multicolumn{1}{c|}{0.625}  & 3.985           & 0.923          & 3.806           & 0.918          & 3.702          & 0.913          \\
                                                               & 40                                  & 0.874                         & 0.924                         & 0.958                         & \multicolumn{1}{c}{0.225}     & \multicolumn{1}{c}{0.375}     & \multicolumn{1}{c|}{0.6}    & 3.668           & 0.878          & 3.561           & 0.864          & 3.496          & 0.858          \\
                                                               & 60                                  & 0.856                         & 0.906                         & 0.942                         & \multicolumn{1}{c}{0.225}     & \multicolumn{1}{c}{0.325}     & \multicolumn{1}{c|}{0.35}   & 3.565           & 0.845          & 3.480           & 0.822          & 3.415          & 0.812          \\
                                                               & 80                                  & 0.835                         & 0.884                         & 0.930                         & \multicolumn{1}{c}{0.2}       & \multicolumn{1}{c}{0.25}      & \multicolumn{1}{c|}{0.325}  & 3.520           & 0.793          & 3.461           & 0.786          & 3.407          & 0.775          \\ \hline
Duck Audio                                                     & 20                                  & 0.921                         & 0.960                         & 0.9817                        & \multicolumn{1}{c}{0.425}     & \multicolumn{1}{c}{0.5}       & \multicolumn{1}{c|}{0.7}    & 4.387           & 0.984          & 4.309           & 0.978          & 4.205          & 0.971          \\
                                                               & 40                                  & 0.919                         & 0.958                         & 0.9814                        & \multicolumn{1}{c}{0.425}     & \multicolumn{1}{c}{0.5}       & \multicolumn{1}{c|}{0.7}    & 4.349           & 0.979          & 4.389           & 0.973          & 4.183          & 0.966          \\
                                                               & 60                                  & 0.918                         & 0.958                         & 0.9812                        & \multicolumn{1}{c}{0.425}     & \multicolumn{1}{c}{0.5}       & \multicolumn{1}{c|}{0.725}  & 4.350           & 0.971          & 4.387           & 0.972          & 4.181          & 0.965          \\
                                                               & 80                                  & 0.917                         & 0.957                         & 0.9811                        & \multicolumn{1}{c}{0.425}     & \multicolumn{1}{c}{0.5}       & \multicolumn{1}{c|}{0.725}  & 4.347           & 0.970          & 4.385           & 0.972          & 4.179          & 0.965          \\ \hline
Echo                                                           & 20                                  & 0.917                         & 0.961                         & 0.980                         & \multicolumn{1}{c}{0.4}       & \multicolumn{1}{c}{0.5}       & \multicolumn{1}{c|}{0.65}   & 2.782           & 0.897          & 2.702           & 0.889          & 2.741          & 0.888          \\
                                                               & 40                                  & 0.916                         & 0.959                         & 0.978                         & \multicolumn{1}{c}{0.4}       & \multicolumn{1}{c}{0.425}     & \multicolumn{1}{c|}{0.675}  & 2.200           & 0.817          & 2.192           & 0.813          & 2.181          & 0.806          \\
                                                               & 60                                  & 0.916                         & 0.957                         & 0.974                         & \multicolumn{1}{c}{0.4}       & \multicolumn{1}{c}{0.425}     & \multicolumn{1}{c|}{0.725}  & 1.904           & 0.742          & 1.878           & 0.737          & 1.867          & 0.739          \\
                                                               & 80                                  & 0.912                         & 0.956                         & 0.971                         & \multicolumn{1}{c}{0.325}     & \multicolumn{1}{c}{0.475}     & \multicolumn{1}{c|}{0.675}  & 1.734           & 0.684          & 1.695           & 0.674          & 1.679          & 0.668          \\ \hline
Highpass                                                       & 20                                  & 0.917                         & 0.958                         & 0.981                         & 0.425                         & 0.5                           & 0.75                        & 3.633           & 0.917          & 3.566           & 0.912          & 3.527          & 0.910          \\
                                                               & 40                                  & 0.914                         & 0.955                         & 0.980                         & 0.425                         & 0.525                         & 0.675                       & 3.346           & 0.860          & 3.313           & 0.855          & 3.246          & 0.849          \\
                                                               & 60                                  & 0.911                         & 0.954                         & 0.979                         & 0.4                           & 0.525                         & 0.625                       & 3.261           & 0.812          & 3.227           & 0.809          & 3.216          & 0.799          \\
                                                               & 80                                  & 0.906                         & 0.950                         & 0.978                         & 0.4                           & 0.525                         & 0.625                       & 3.291           & 0.771          & 3.239           & 0.767          & 3.206          & 0.763          \\ \hline
Lowpass                                                        & 20                                  & 0.914                         & 0.953                         & 0.975                         & 0.375                         & 0.475                         & 0.675                       & 4.091           & 0.967          & 4.006           & 0.961          & 3.920          & 0.956          \\
                                                               & 40                                  & 0.896                         & 0.945                         & 0.970                         & 0.3                           & 0.475                         & 0.575                       & 3.999           & 0.945          & 3.915           & 0.940          & 3.833          & 0.936          \\
                                                               & 60                                  & 0.892                         & 0.936                         & 0.966                         & 0.275                         & 0.425                         & 0.55                        & 3.972           & 0.927          & 3.903           & 0.924          & 3.812          & 0.921          \\
                                                               & 80                                  & 0.879                         & 0.931                         & 0.961                         & 0.25                          & 0.425                         & 0.55                        & 3.978           & 0.917          & 3.908           & 0.914          & 3.843          & 0.910          \\ \hline
Pink Noise                                                     & 20                                  & 0.924                         & 0.961                         & 0.9816                        & 0.375                         & 0.5                           & 0.7                         & 4.269           & 0.978          & 4.229           & 0.972          & 4.151          & 0.967          \\
                                                               & 40                                  & 0.924                         & 0.960                         & 0.9812                        & 0.35                          & 0.475                         & 0.7                         & 4.186           & 0.970          & 4.164           & 0.965          & 4.094          & 0.959          \\
                                                               & 60                                  & 0.924                         & 0.962                         & 0.9802                        & 0.35                          & 0.475                         & 0.7                         & 4.135           & 0.965          & 4.119           & 0.960          & 4.054          & 0.955          \\
                                                               & 80                                  & 0.924                         & 0.962                         & 0.980                         & 0.375                         & 0.5                           & 0.675                       & 4.103           & 0.960          & 4.086           & 0.956          & 4.025          & 0.951          \\ \hline
White Noise                                                    & 20                                  & 0.881                         & 0.925                         & 0.958                         & 0.1                           & 0.275                         & 0.325                       & 3.310           & 0.926          & 3.311           & 0.922          & 3.292          & 0.918          \\
                                                               & 40                                  & 0.839                         & 0.896                         & 0.934                         & 0.025                         & 0.05                          & 0.175                       & 2.996           & 0.902          & 3.022           & 0.899          & 2.991          & 0.897          \\
                                                               & 60                                  & 0.799                         & 0.864                         & 0.902                         & 0                             & 0.05                          & 0.05                        & 2.844           & 0.887          & 2.867           & 0.886          & 2.842          & 0.884          \\
                                                               & 80                                  & 0.762                         & 0.833                         & 0.889                         & 0                             & 0                             & 0.025                       & 2.768           & 0.881          & 2.801           & 0.882          & 2.773          & 0.880          \\ \hline
\end{tabular}
\end{table*}

\section{Sync Pattern Detection Accuracy}
\label{sec:appendix-sync-patterns}

\begin{figure}[!h]
    \centering
    \includegraphics[width=1\linewidth]{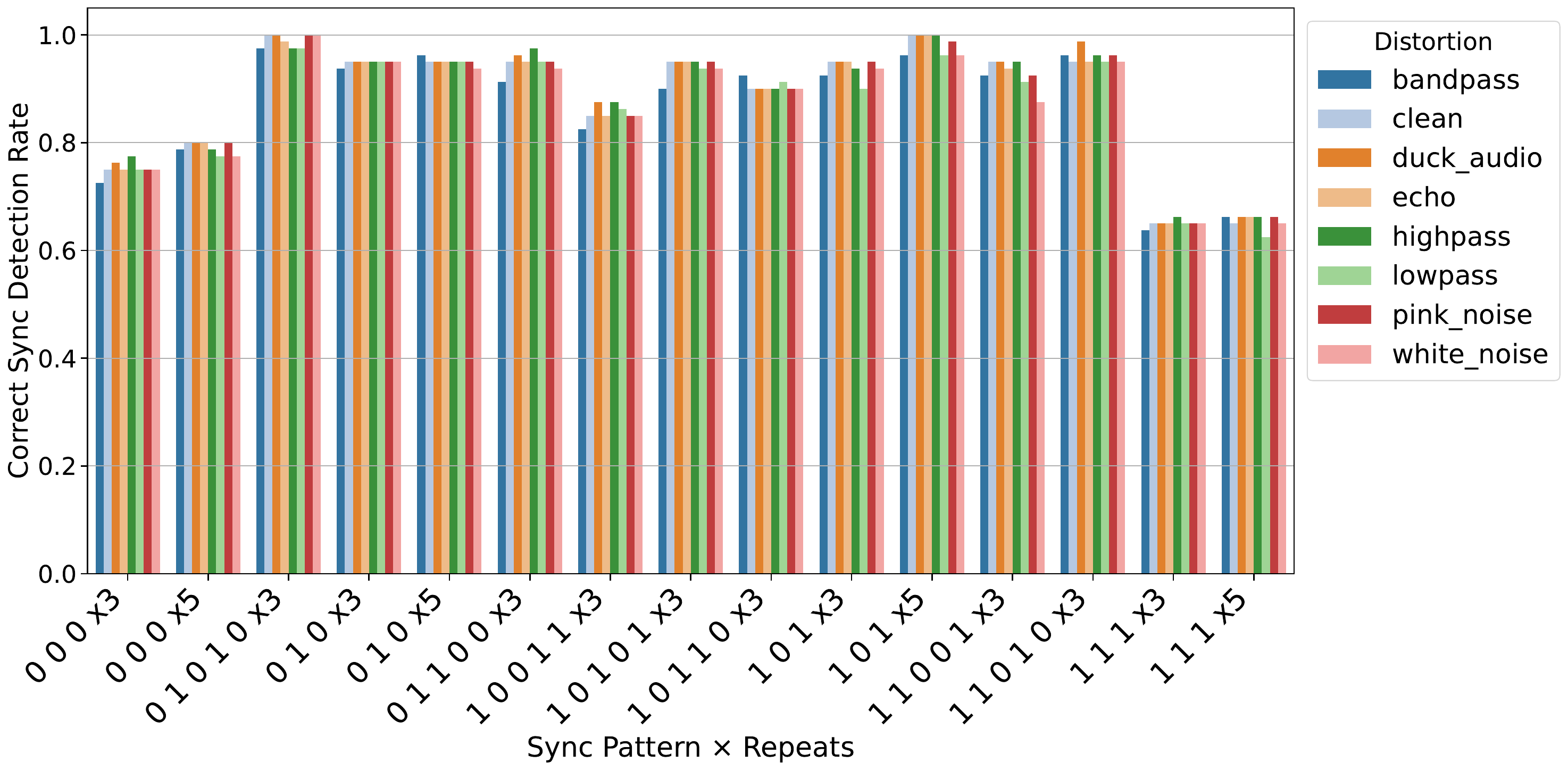}
    \caption{Sync detection accuracy across 15 sync patterns under varying distortions.}
    \label{fig:sync-pattern-comparison}
\end{figure}

Figure~\ref{fig:sync-pattern-comparison} shows the full breakdown of sync detection accuracy across all 15 evaluated patterns under different distortion types. Full per-distortion breakdowns appear in Appendix~\ref{sec:appendix-sync-patterns}.
These results complement Table~\ref{tab:sync-all} in the main text and illustrate per-distortion variation, highlighting the stability of high-entropy patterns such as \texttt{[0,1,0,1,0]} and the fragility of uniform sequences like \texttt{[0,0,0]} and \texttt{[1,1,1]}.

\section{Additional Protocol Stage Results}
\label{sec:appendix-protocol-plots}

\begin{figure}[!h]
    \centering
    \begin{subfigure}[t]{0.48\textwidth}
        \includegraphics[width=\linewidth]{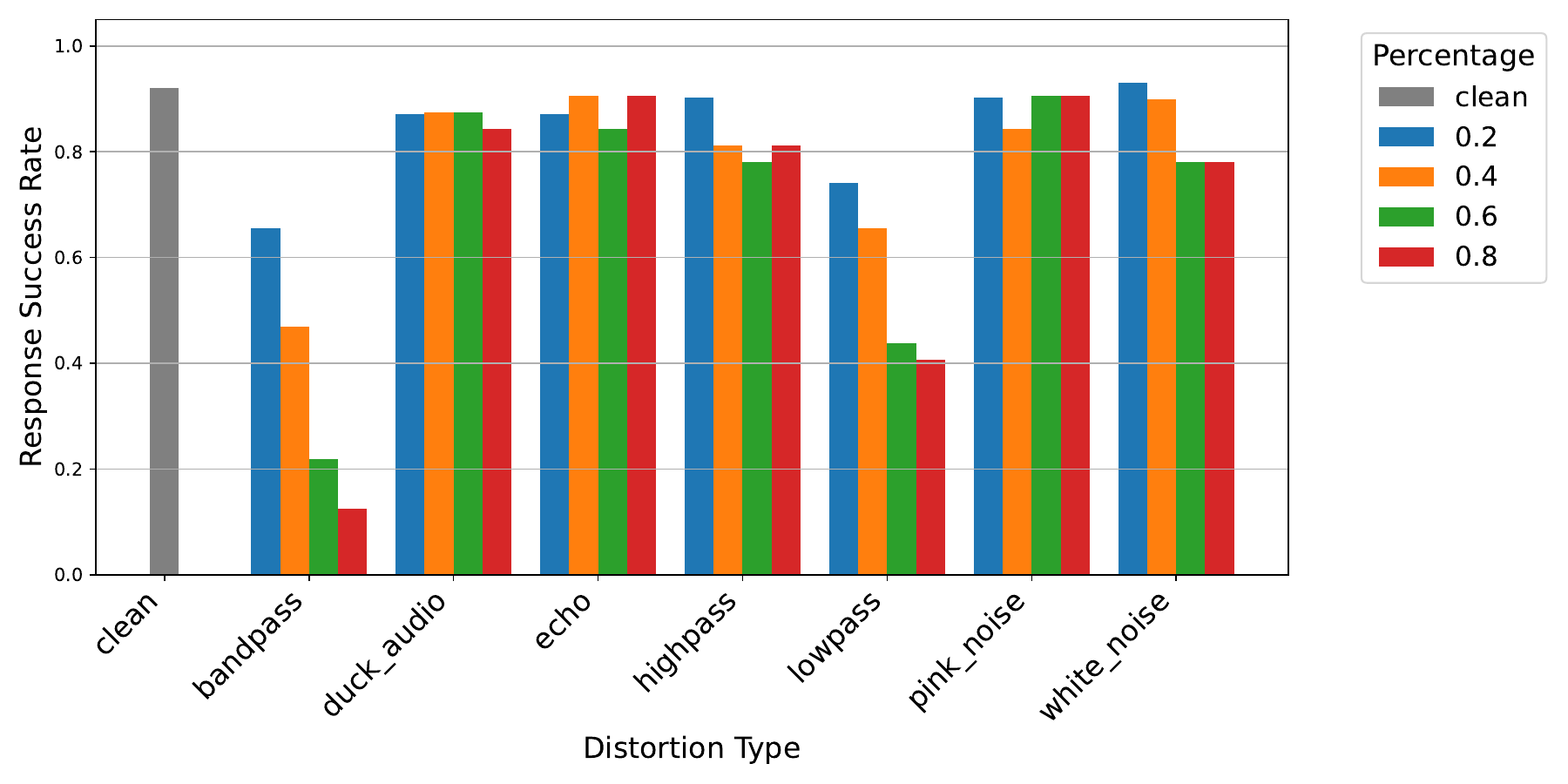}
        \caption{Response Success Rate}
    \end{subfigure}
    \hfill
    \begin{subfigure}[t]{0.48\textwidth}
        \includegraphics[width=\linewidth]{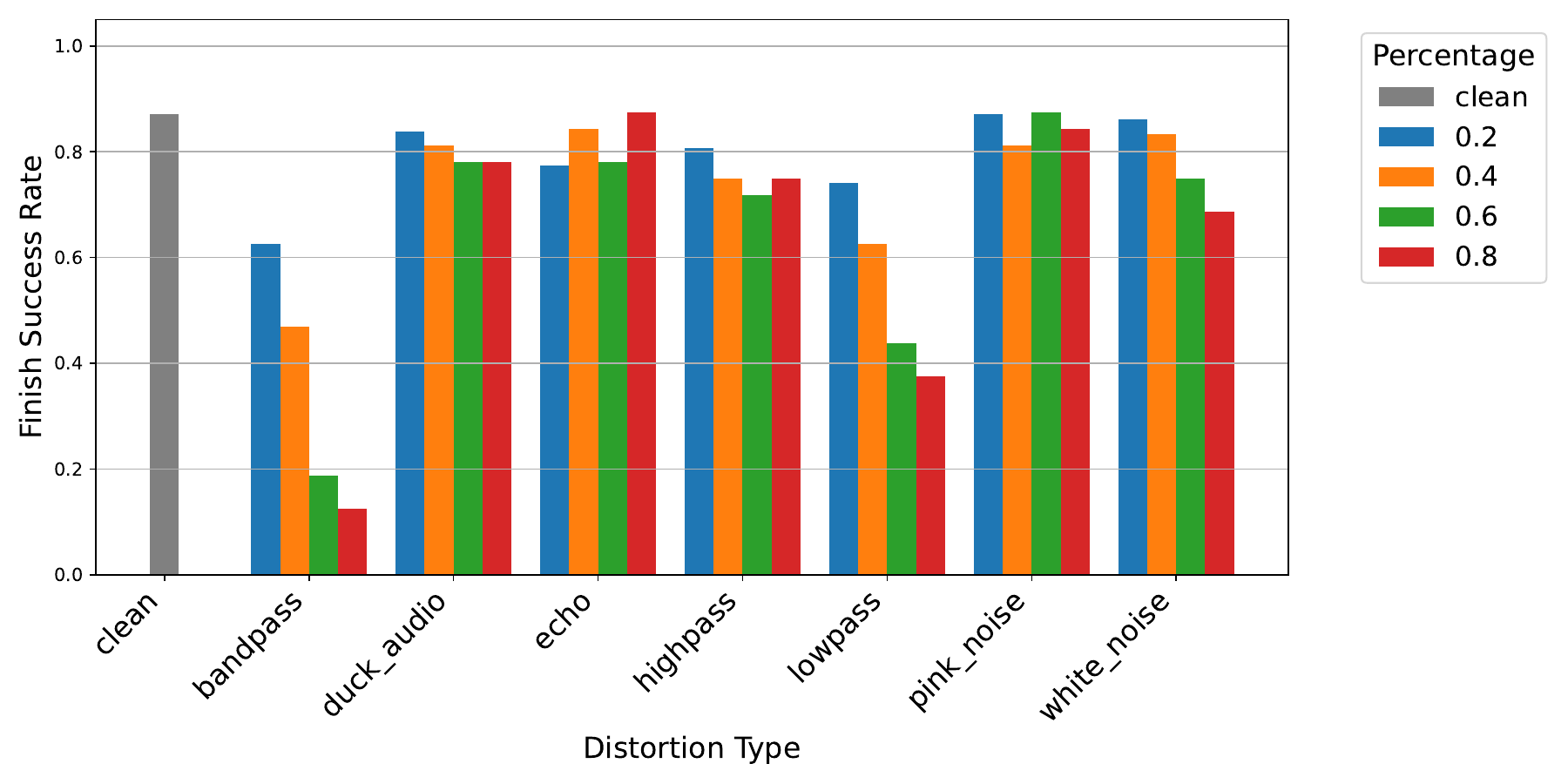}
        \caption{Finish Success Rate}
    \end{subfigure}
    \caption{Success rates for the response and finish stages across distortion types and coverage levels. These complement the main protocol evaluation in Figure~\ref{fig:protocol-success-main}.}
    \label{fig:per-distortion-plots}
\end{figure}

Figure~\ref{fig:per-distortion-plots} provides the full per-stage breakdown for the \emph{response} and \emph{finish} phases of the protocol. These stages were omitted from the main text for brevity but follow similar trends: performance degrades under bursty or midband-eliminating distortions (e.g., white noise, echo, bandpass), and remains stable under milder effects (e.g., pink noise, ducking).

The response phase is particularly sensitive to bursty distortions, which corrupt the longer MAC payload. The finish beacon, while structurally similar to the start beacon, is transmitted only after a valid response, and thus inherits cumulative failure probabilities. As shown, its performance closely mirrors that of the response stage across distortion types and levels.

\section{ Authentication Failures}
\label{sec:appendix-failures}
\begin{figure}[t]
  \centering
  \includegraphics[width=\linewidth]{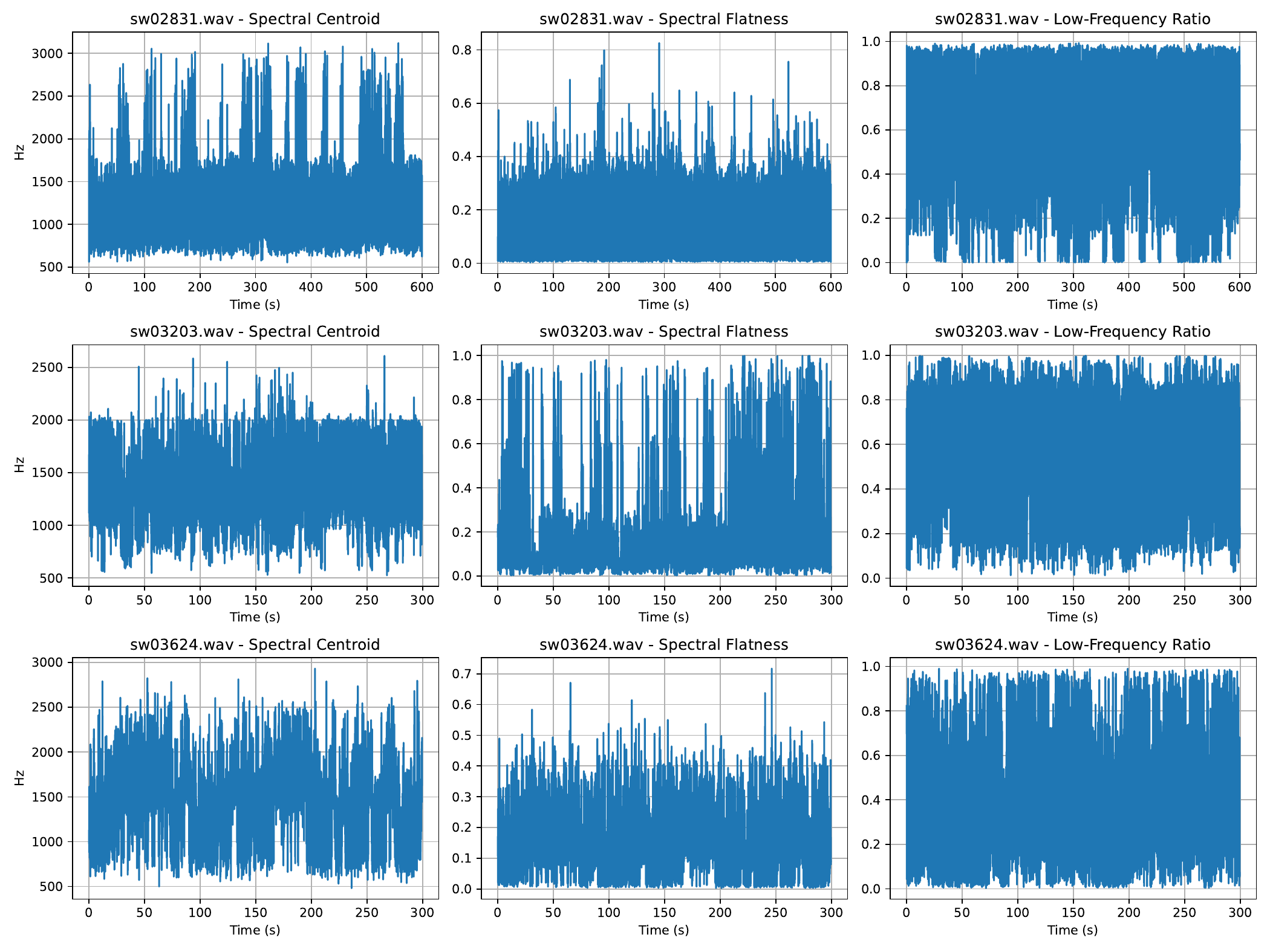}
  \caption{Spectral feature evolution over time for retry-1 success cases. Each row shows 60 seconds of audio. Segments with low centroid, low flatness, and high low-frequency energy ratio correspond to acoustically sparse or tonal conditions where watermark decoding is likely to fail.}
  \label{fig:feature-drift}
\end{figure}

To better understand why authentication fails at low watermark strengths, we analyze spectral properties over time for audio files that succeed only upon retry. Figure~\ref{fig:feature-drift} shows the frame-level evolution of spectral centroid, flatness, and low-frequency energy ratio for three such examples. In failure-prone segments (e.g., \texttt{sw02831}), the spectral centroid remains low ($<$600\,Hz), spectral flatness is minimal, and over 80\% of energy lies below 500\,Hz. These conditions indicate voiced, tonal content with little high-frequency energy, which restricts watermark embedding capacity at $\alpha=0.6$. Other files (e.g., \texttt{sw03624}) exhibit greater spectral diversity, with centroid excursions above 1\,kHz and increased flatness in later segments—likely corresponding to unvoiced consonants or fricatives. These richer spectral regions enable watermark recovery at higher $\alpha$, validating our hypothesis that spectral sparsity and low pitch inhibit decoder robustness at low energy levels.

\section{Bit Recovery Improvements at Higher Alpha}
\label{sec:appendix-bit-corrections}

We present spectrogram visualizations showing representative 40 ms segments where the embedded bit was not correctly decoded at $\alpha = 0.6$, but was successfully recovered when the watermark strength was increased to $\alpha = 0.8$. These examples illustrate how stronger watermarks yield clearer spectral structures that are more easily detected by the decoder. In particular, the $\alpha = 0.8$ spectrograms exhibit more pronounced energy patterns — often forming vertical striations or frequency-localized bursts — which contrast with the more diffuse and low-energy perturbations seen at $\alpha = 0.6$. These structural differences make the watermark more distinguishable from the host audio content, especially in mid-frequency regions where perceptual masking is less dominant. By visually comparing the decoded failures at $\alpha = 0.6$ with the corrected cases at $\alpha = 0.8$, we confirm that increasing $\alpha$ improves the reliability of the embedded bits.
We visualize representative 40\,ms spectrogram segments where decoding failed at watermark strength $\alpha = 0.6$ but succeeded at $\alpha = 0.8$. These examples highlight how increasing $\alpha$ improves watermark detectability. At $\alpha = 0.6$, the perturbations tend to be diffuse and low-energy, making them harder for the decoder to extract reliably. In contrast, the $\alpha = 0.8$ spectrograms exhibit clearer vertical patterns and more localized spectral energy—particularly in mid-frequency bands—leading to more confident and accurate bit recovery.

\begin{figure}[]
    \centering
    \begin{subfigure}[b]{0.49\linewidth}
        \includegraphics[width=\linewidth]{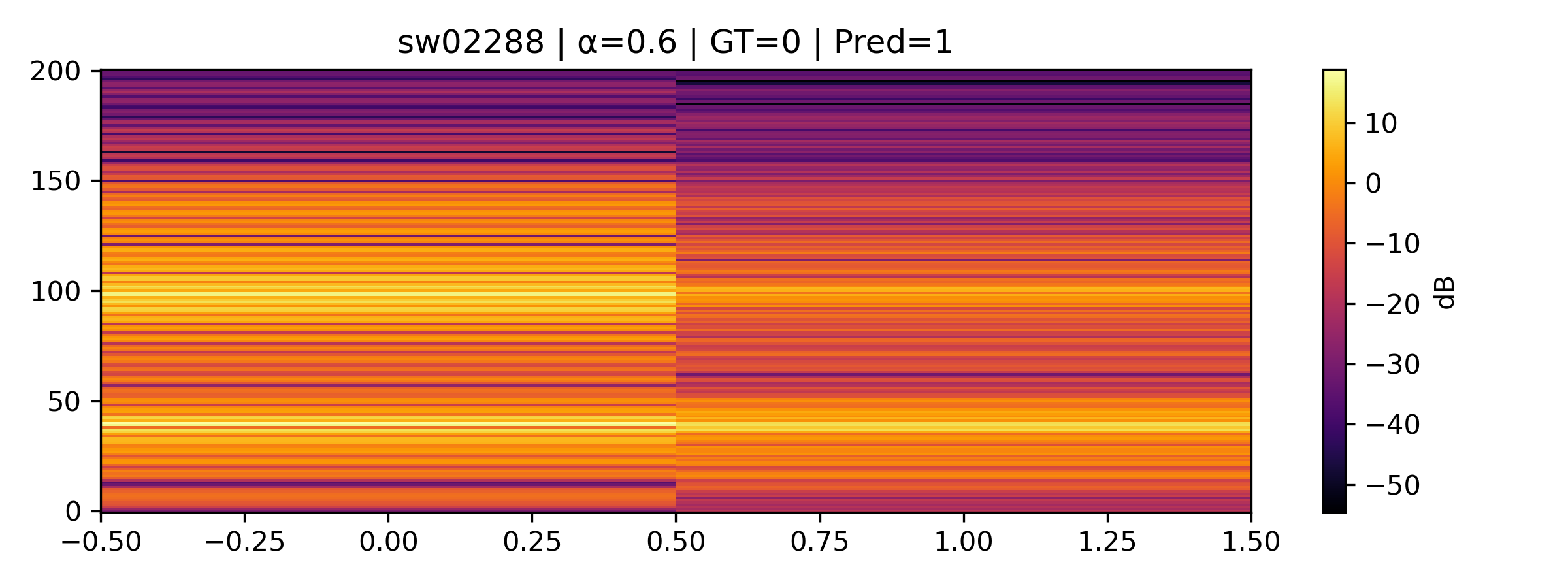}
        \caption*{\tiny $\alpha{=}0.6$\\GT=0, Pred=1}
    \end{subfigure}
    \hfill
    \begin{subfigure}[b]{0.49\linewidth}
        \includegraphics[width=\linewidth]{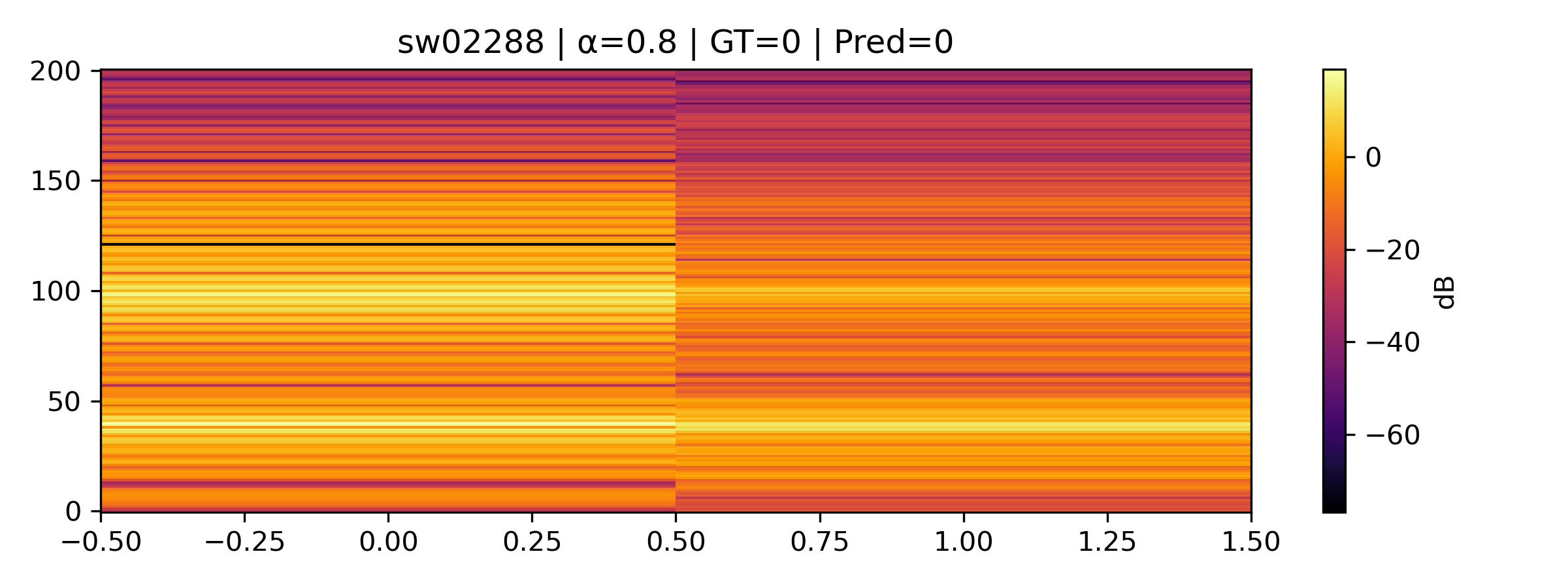}
        \caption*{\tiny $\alpha{=}0.8$\\GT=0, Pred=0}
    \end{subfigure}

    \begin{subfigure}[b]{0.49\linewidth}
        \includegraphics[width=\linewidth]{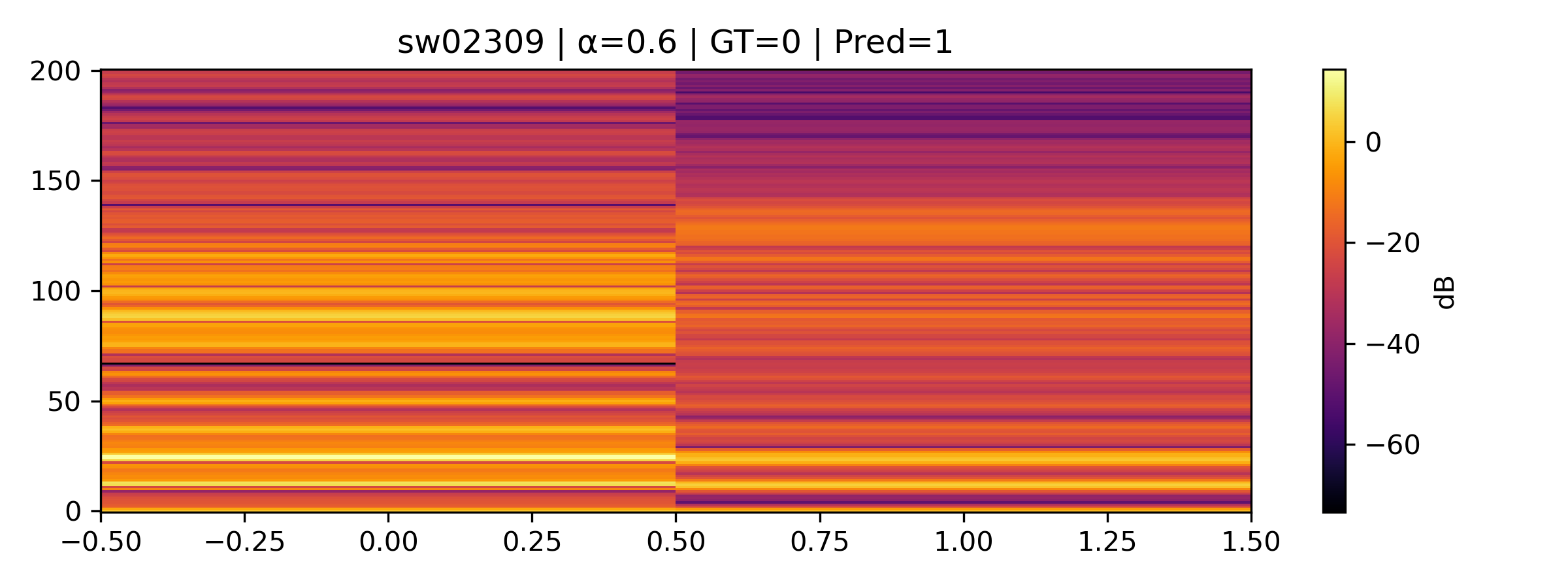}
        \caption*{\tiny $\alpha{=}0.6$\\GT=0, Pred=1}
    \end{subfigure}
    \hfill
    \begin{subfigure}[b]{0.49\linewidth}
        \includegraphics[width=\linewidth]{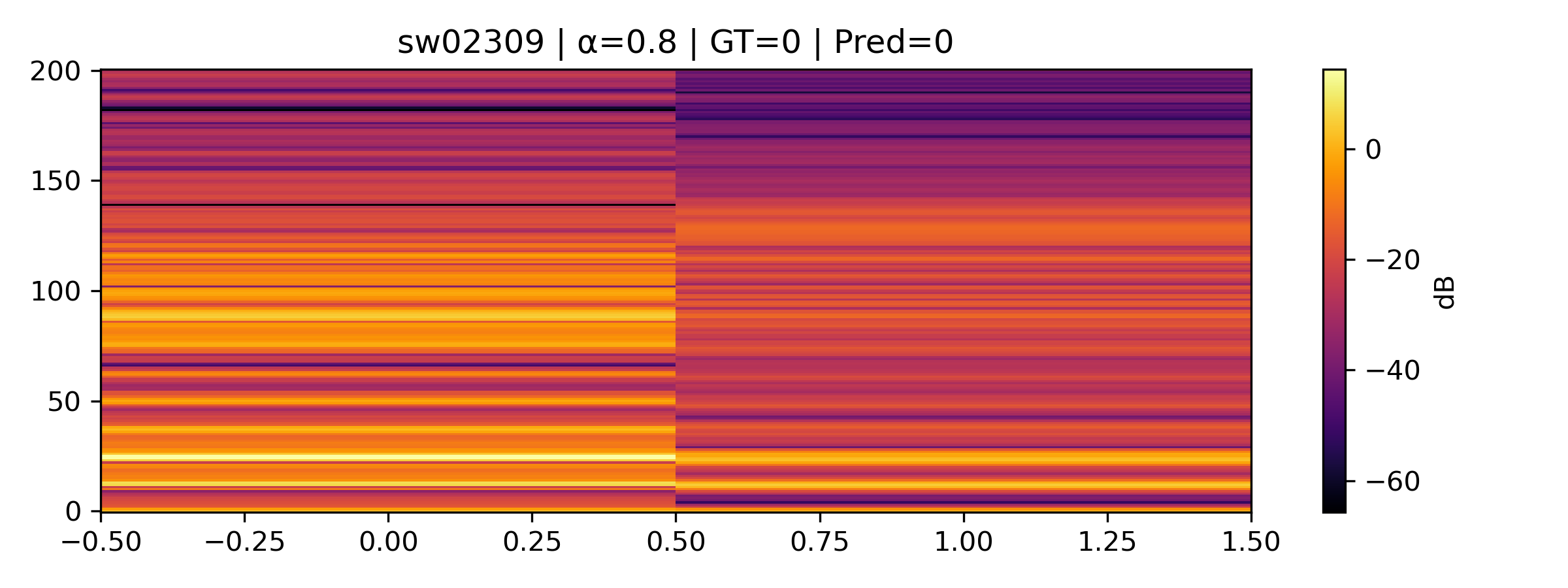}
        \caption*{\tiny $\alpha{=}0.8$\\GT=0, Pred=0}
    \end{subfigure}
    
    \vspace{1mm}
    
    \begin{subfigure}[b]{0.49\linewidth}
        \includegraphics[width=\linewidth]{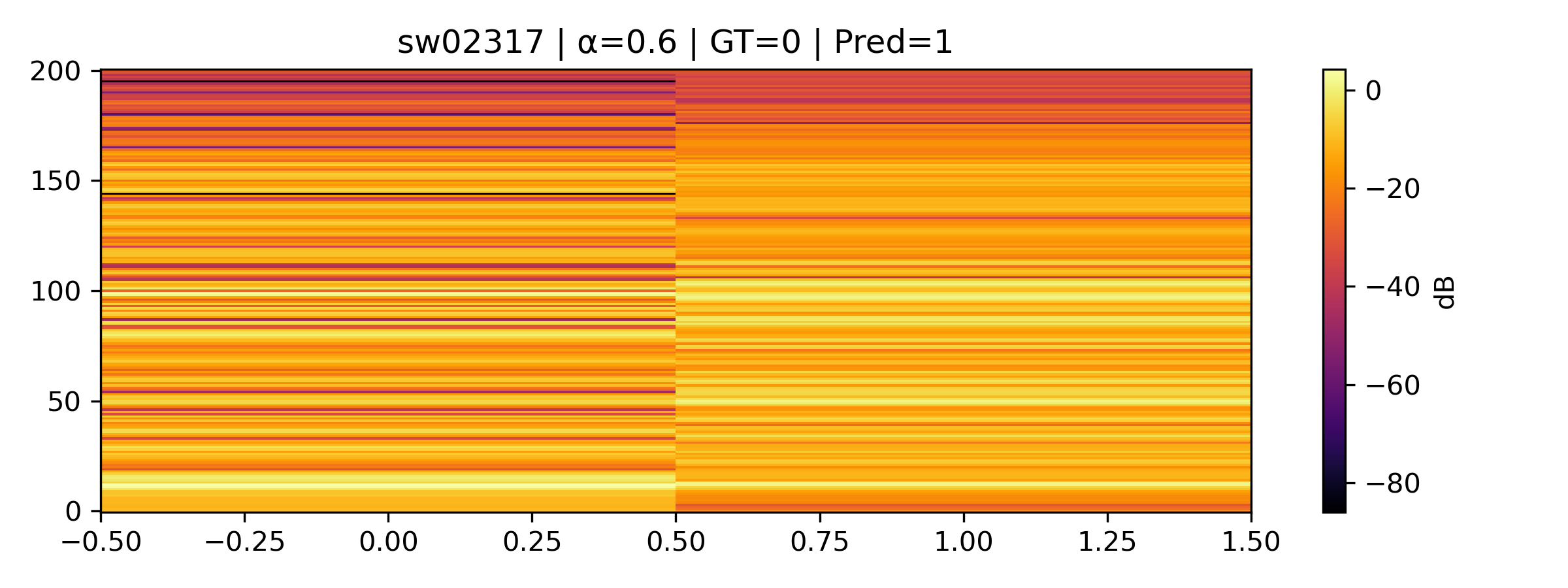}
        \caption*{\tiny $\alpha{=}0.6$\\GT=0, Pred=1}
    \end{subfigure}
    \hfill
    \begin{subfigure}[b]{0.49\linewidth}
        \includegraphics[width=\linewidth]{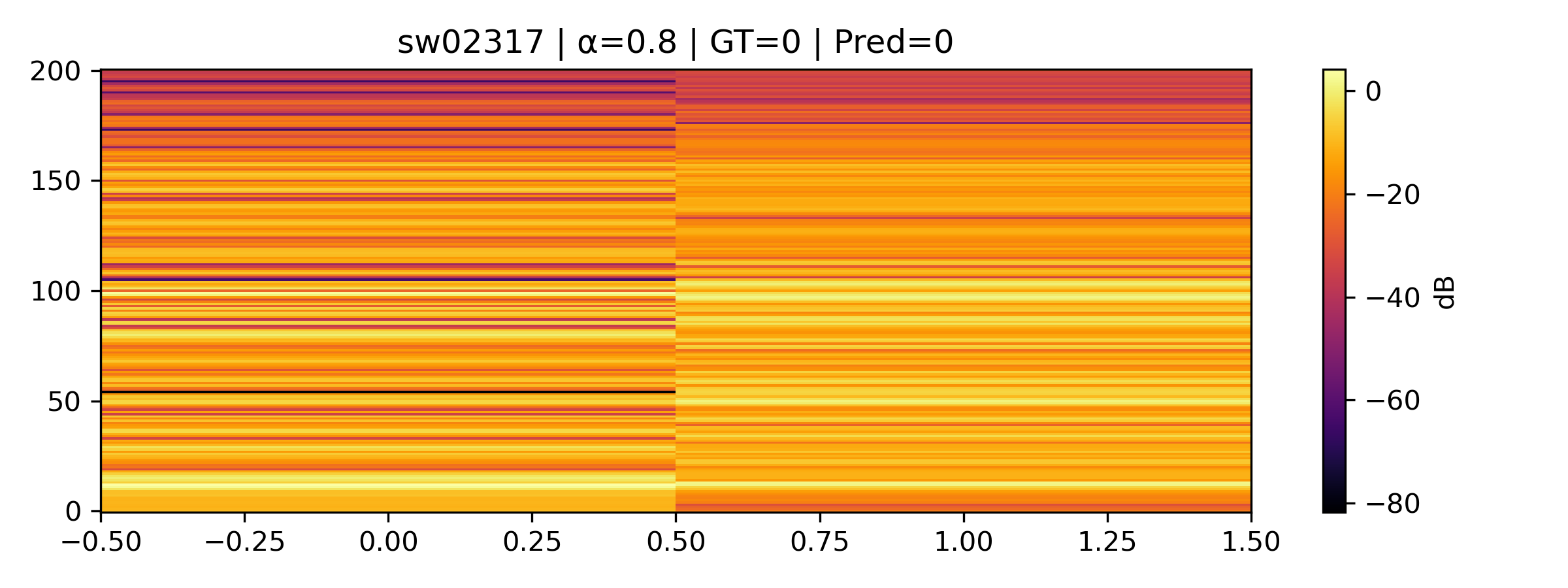}
        \caption*{\tiny $\alpha{=}0.8$\\GT=0, Pred=0}
    \end{subfigure}

    \begin{subfigure}[b]{0.49\linewidth}
        \includegraphics[width=\linewidth]{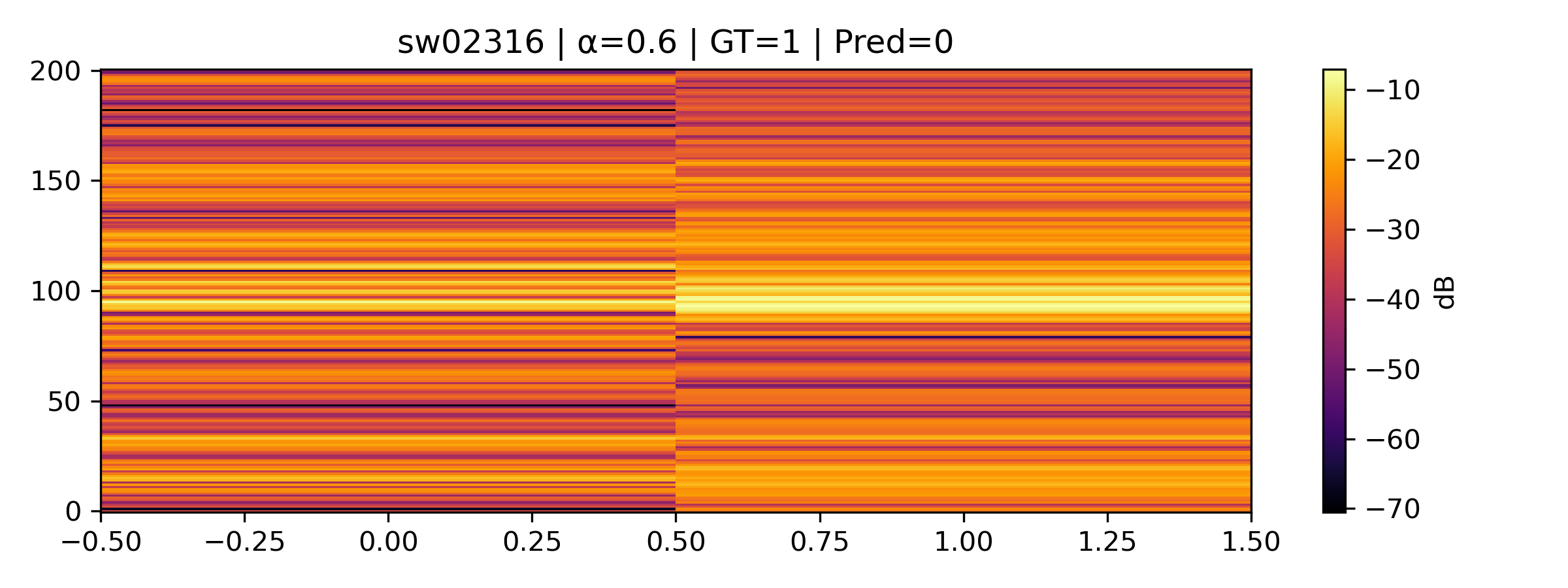}
        \caption*{\tiny $\alpha{=}0.6$\\GT=1, Pred=0}
    \end{subfigure}
    \hfill
    \begin{subfigure}[b]{0.49\linewidth}
        \includegraphics[width=\linewidth]{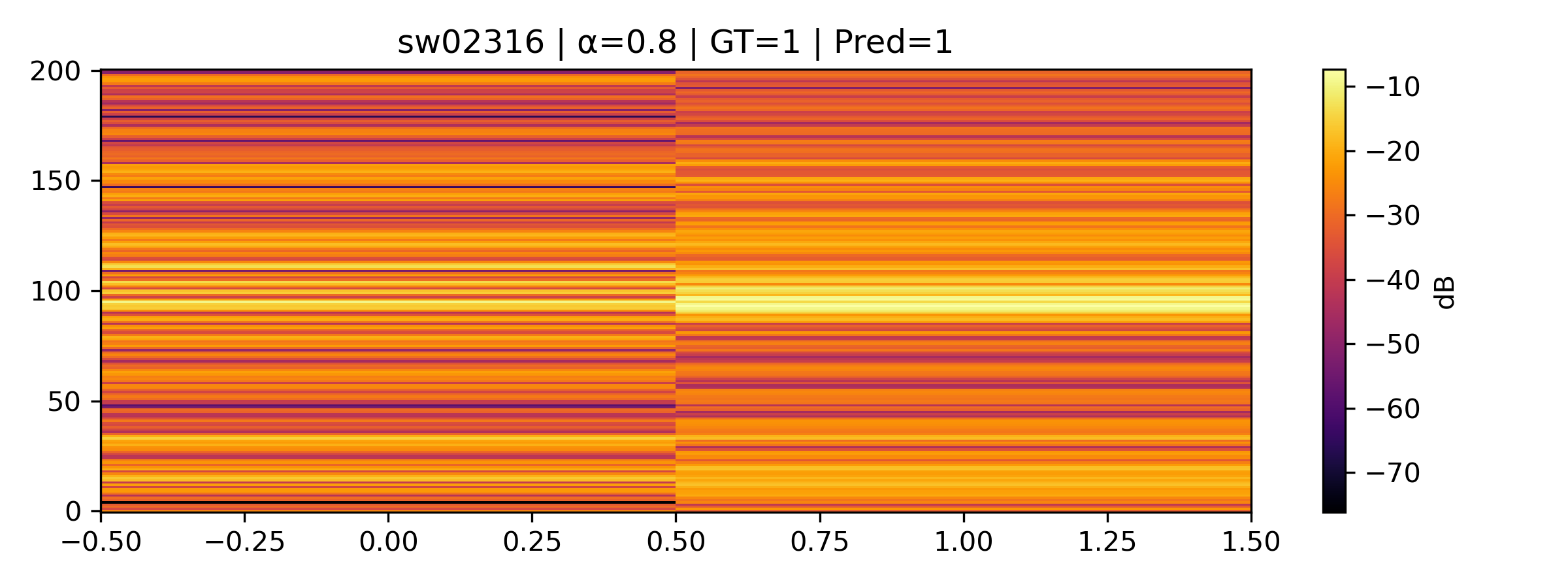}
        \caption*{\tiny $\alpha{=}0.8$\\GT=1, Pred=1}
    \end{subfigure}
    
    \caption{Spectrograms of several 40\,ms audio segments where decoding failed at $\alpha = 0.6$ but succeeded at $\alpha = 0.8$. Increased watermark strength produces more structured spectral energy, improving decoder accuracy.}
    \label{fig:alpha-spectrograms}
\end{figure}

\end{document}